\documentclass[sn-mathphys-num]{sn-jnl}

\usepackage[T1]{fontenc}
\usepackage{graphicx}
\usepackage{multirow}
\usepackage{amsmath,amssymb,amsfonts}
\usepackage{amsthm}
\usepackage{mathrsfs}
\usepackage[title]{appendix}
\usepackage{xcolor}
\usepackage{textcomp}
\usepackage{manyfoot}
\usepackage{booktabs}
\usepackage{algorithm}
\usepackage{algorithmicx}
\usepackage{algpseudocode}
\usepackage{listings}
\usepackage{upgreek}
\usepackage[absolute]{textpos}
\usepackage[export]{adjustbox}
\usepackage{float}
\usepackage{dsfont}
\usepackage{cleveref}

\usepackage{siunitx}
\DeclareSIUnit\year{a}
\DeclareSIUnit\month{mon}
\DeclareSIUnit\day{d}
\DeclareSIUnit\hour{h}
\DeclareSIUnit\minute{min}

\usepackage{soul}
\usepackage{ulem}

\newcommand{\upA}{\mathrm{A}}
\newcommand{\upB}{\mathrm{B}}

\newcommand{\upDL}{\mathrm{DL}}
\newcommand{\upUL}{\mathrm{UL}}
\newcommand{\upIS}{\mathrm{IS}}

\newcommand{\BSM}{\mathrm{BSM}}

\newcommand{\upc}{\mathrm{c}}
\newcommand{\upd}{\mathrm{d}}
\newcommand{\upe}{\mathrm{e}}

\newcommand{\Tx}{\text{Tx}}
\newcommand{\Rx}{\text{Rx}}
\newcommand{\GS}{\text{GS}}
\newcommand{\sat}{\text{sat}}

\begin{document}

\title{Simulation of satellite and optical link dynamics in a quantum repeater constellation}

\author*[1]{\fnm{Jaspar} \sur{Meister}}\email{jaspar.meister@dlr.de}

\author[2]{\fnm{Philipp} \sur{Kleinpa\ss}}

\author[2]{\fnm{Davide} \sur{Orsucci}}

\affil*[1]{\orgdiv{Institute for Satellite Geodesy and Inertial Sensing}, \orgname{German Aerospace Center (DLR)}, \orgaddress{\street{Callinstr. 30b}, \city{Hanover}, \postcode{30167}, \country{Germany}}}

\affil[2]{\orgdiv{Institute of Communications and Navigation}, \orgname{German Aerospace Center (DLR)}, \orgaddress{\street{Münchener Str. 20}, \city{We\ss ling}, \postcode{82234}, \country{Germany}}}

\abstract{
Quantum repeaters and satellite-based optical links are complementary technological approaches to overcome the exponential photon loss in optical fibers and thus allow quantum communication on a global scale. We analyze architectures which combine these approaches and use satellites as quantum repeater nodes to distribute entanglement to distant optical ground stations. Here we simulate dynamic, three-dimensional ground station passes, going beyond previous studies that typically consider static satellite links. For this, we numerically solve the equations of motion of the dynamic system consisting of three satellites in low Earth orbit. The model of the optical link takes into account atmospheric attenuation, single-mode fiber coupling, beam wandering and broadening, as well as adaptive optics effects. We derive analytical expressions for the Bell state measurement and associated error rates for quantum memory assisted communications, including retrieval efficiency and state coherence. We consider downlink and uplink architectures for continental and intercontinental connections and evaluate the impact of satellite altitude and inter-satellite distance on the expected entanglement swapping rate. Our simulation model enables us to design different orbital configurations for the satellite constellation and analyze the annual performance of the quantum repeater under realistic conditions.}


\maketitle

\section{Introduction}\label{sec:intro}

Quantum networks could enable applications such as distributed quantum computation~\cite{caleffi2022distributed} and sensing~\cite{zhang2021distributed}, high-accuracy clock synchronisation~\cite{jozsa2000quantum} and secure communications through the use of Quantum Key Distribution (QKD)~\cite{xu2020secure}. If distances of a few hundred kilometers or less need to be bridged, quantum networks based on optical fibers can be employed~\cite{Boaron2018}. However, when envisioning quantum links on a continental or even global scale, direct fiber connections between endpoints are ineffective, since communication losses increase exponentially with the link distance.

In classical communications, a signal can be amplified at regular intervals along the fiber to compensate for losses and reach greater distances. In quantum communications, the noiseless amplification of a signal is prohibited by the no-cloning theorem~\cite{wootters1982single}. To increase the communication distance, Briegel, Dür, Cirac and Zoller~\cite{Briegel1998} introduced the concept of a Quantum Repeater (QR). It uses entangled photons and purification protocols to create high-fidelity entanglement over long distances and then allows the transmission of quantum information via quantum teleportation protocols. In the following years, this has triggered many theoretical and experimental developments for fiber-based quantum repeater networks, see~\cite{Azuma2022} for a recent review. 

In theory, the interleaving of sufficient numbers of quantum repeaters would enable quantum networks to achieve high entanglement distribution rates over arbitrarily long distances. The photonic states are subject to noise caused by device imperfections, which increases with the number of intermediate repeater nodes until the transmission channel becomes entanglement-breaking~\cite{horodecki2003}. Ultimately, the noise must be reduced by using entanglement purification schemes that combine several noisy entangled pairs to produce fewer high-fidelity entangled pairs~\cite{purification}. However, this approach would require hundreds of intermediate repeater nodes, each hosting a full-fledged quantum computer that performs entanglement purification to bridge intercontinental distances~\cite{azuma2016}. 

Satellite-based links can be advantageous for quantum communication networks because they are only subject to loss that scales quadratically with distance due to beam divergence, rather than exponential loss due to optical attenuation in fibers. This is a technological approach that is already available today, as demonstrated by recent developments in both the theoretical~\cite{Ecker2021, Khatri2021, Roger2023, Islam2024} and experimental implementation of satellite-based QKD~\cite{liao2017, liao2018, Chen2021, yin2020}. In particular, the first satellite QKD downlink experiment was conducted with the Micius satellite in 2017~\cite{liao2017} and later integrated with a fiber-based QKD network on the ground~\cite{Chen2021}.

If the key is only established between a satellite and an optical ground station, it is of little use, as the key should usually be exchanged between communication parties at different locations on Earth. Therefore, the most common approach to connect them is to use the satellite as a trusted node. Using a prepare-and-measure protocol, the satellite establishes a key with both participants individually, after which they can agree on a shared key. This key is then known to them and the satellite. To prevent the satellite from gaining knowledge of the final key, a source-independent protocol may be employed, e.g. the entanglement-based BBM92 protocol~\cite{BBM92}. Although it provides a higher level of security, a direct implementation requires a simultaneous line-of-sight between the satellite and both ground stations as well as the successful transmission of two entangled photons in each round. This leads to a scaling of the key rate with the product of the individual channel transmissions and thus a significantly worse performance compared to a trusted-node implementation. To improve the rate of the entanglement distribution, satellite and QR technologies can be combined, which would allow quantum communication links to be established over long distances with only a few repeater nodes. A single-node QR link may be employed, for which several architectures have been proposed, such as \textit{meet-in-the-middle}, \textit{sender-receiver} and \textit{midpoint-source}~\cite{Jones2016}.

However, quantum communication via satellite poses other challenges. One major challenge arises from the fact that the connection time from satellite to a ground station is usually short, depending on the satellite orbit. For Low Earth Orbits (LEO) with an altitude of \SI{500}{\kilo\meter}, the simultaneous connection to both ground station is only a few minutes per day. The high background radiation from sunlight limits operation at ground station sites to nighttime, unless very strong spectral, spatial and temporal filtering is applied~\cite{gruneisen2021}. In addition, clouds and fog can prevent the communication over optical wavelengths altogether. The high volatility of the communication rate renders it difficult to assess the performance of satellite-based quantum repeaters. Therefore, annual averages must be determined in order to meaningfully estimate communication rates.

Compared to fiber-based quantum networks that can be tested in laboratory environments, testing space-based technologies in orbit is associated with high costs and long preparation times. Consequently, simulation models offer a cost-effective alternative for the early design phase to investigate different satellite-based QR architectures. Several software tools have been developed for the simulation of quantum networks, including \textit{SimulaQron}~\cite{simulaqron}, \textit{QuISP}~\cite{quisp}, \textit{NetSquid}~\cite{netsquid}, \textit{SeQUeNCe}~\cite{sequence} and \textit{ReQuSim}~\cite{requsim}. These tools are primarily event-based systems designed to simulate each qubit generated and processed in the network, including the error processes that affect them and the necessary classical support systems. This approach enables sophisticated Monte-Carlo analyses of network performance, taking into account the limitations and imperfections of real systems. However, these tools are tailored for fiber-based systems, which are characterized by static link and node configurations, while satellite-based repeaters experience constantly changing link parameters. Therefore, the instantaneous amount of entanglement swaps has to be evaluated over the entire satellite pass with a resolution of a few seconds. It is possible to estimate the number of entanglement swaps at each point in time via Monte-Carlo sampling, as recently shown in ref.~\cite{fittipaldi2024entanglement}. However, this is computationally expensive, especially as multiple evaluations are necessary to optimize the free parameters of the repeater protocol, e.g. the cutoff time of the maximum memory storage.

This work proposes the application of a satellite orbit propagator to simulate the satellite position and velocity vector while analytically evaluating the number of entanglement swaps at each simulation time step. Our simulation model computes the performance for optical Uplink (UL) and Downlink (DL) channels of a QR consisting of three satellites, one of which contains Quantum Memories (QM). We take into account the rotation of Earth, shadow conditions, and the link dynamics of the three-dimensional ground station pass of a satellite. By fully integrating our QR simulation with our orbit propagator, we are able to analyze arbitrary orbits for QRs and simulate the quantum link for periods of up to one year, including effects arising from the non-spherical potential of Earth. This allows for greater flexibility in analyzing possible QR operating scenarios and architectures. 

In \cref{sec:QRA} we describe our two satellite-based QR architectures, including the entanglement swapping procedure and the required satellite communication link. \Cref{sec:sat} provides background information on satellite dynamics. In \cref{sec:performance} we show our results of the QR performance analysis, starting with single zenith passes of the satellites, followed by more general ground station passes and long-term studies. \Cref{sec:dis} presents a critical discussion and our conclusions.

\section{Satellite-based quantum repeaters}\label{sec:QRA}

\begin{figure}
    \centering
    \includegraphics[width=\textwidth]{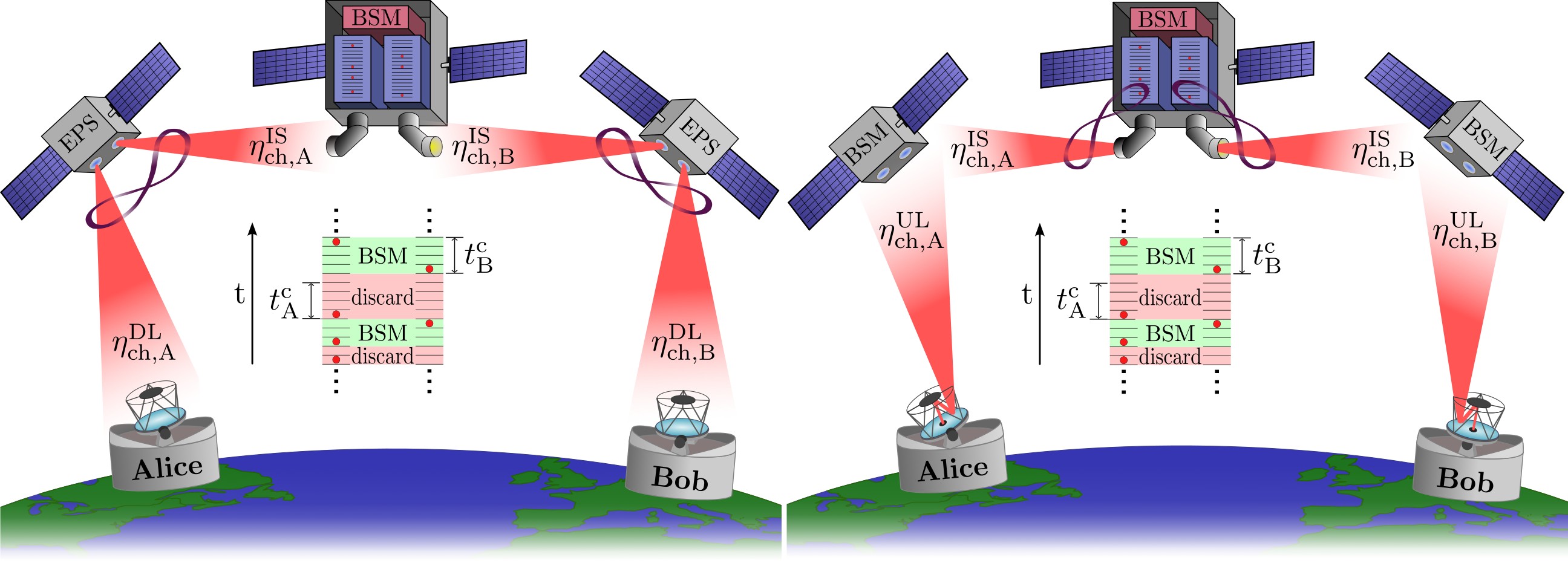} 
    \caption{Illustration of the two reference QR architectures with the downlink (DL) architecture on the left and the uplink (UL) architecture on the right. In both scenarios, two QMs, labeled A and B, are mounted on the central satellite, which store quantum states up to some cutoff times $t_\upA^\upc, t_\upB^\upc$, after which the states are discarded. This configuration allows Bell state measurements to be performed between two asynchronously stored quantum states, one in memory A and one in B.
    In the DL architecture, entangled photon pairs are generated in the outer satellites and transmitted to the ground stations and to the central satellite with link efficiencies $\eta_{\text{ch},i}^\upDL$ and $\eta_{\text{ch},i}^\upIS$, for $i \in \{\text{A},\text{B}\}$ respectively. In the central satellite, an explicit heralding mechanism is required to indicate the arrival of a photon in a QM. Additionally, a classical signal confirming the arrival of its entangled partner at the respective ground station needs to be transmitted to the central satellite. Only then can the photon be used for an entanglement swap. Since there is usually no direct line-of-sight between the ground stations and the central satellite, the classical signals are routed via the outer satellites. The corresponding signal propagation times are indicated as $t_\GS$ and $t_\sat$.
    In the UL architecture, the propagation directions of the quantum signals are reversed, i.e. the ground stations perform an uplink to the outer satellites with channel efficiencies $\eta_{\text{ch},i}^\upUL$. The central satellite generates entangled memory-photon states and sends them to the outer satellites, which perform Bell state measurement on simultaneously arriving photons. On the bottom a representation of the pairing strategy for the entanglement swapping procedure is displayed.}
    \label{QR_Architecture}
\end{figure}

The QR scheme considered in this work consists of a single satellite-based repeater node and two outer support satellites. The reason for using a single repeater node is as follows. Quantum state processing is inherently noisy and more errors accumulate when more intermediate nodes are present, which eventually results in the complete loss of the transmitted quantum information~\cite{horodecki2003}. In principle, these errors can be corrected using entanglement purification protocols~\cite{purification}, but these require the use of full-fledged quantum computers and are therefore not viable as a short-term or mid-term solution~\cite{azuma2016}. In contrast, a single repeater node link could allow end-to-end entanglement distribution with limited amount of noise and without the need for entanglement purification. Therefore, it is more likely that the resulting scheme can be implemented with near-future technologies.

The general aim of the scheme is to distribute entanglement between users who are located far away from each other on the ground. Otherwise, the scheme is application-agnostic: some possibilities include establishing distributed quantum computation~\cite{distributedQC}, quantum sensor networks~\cite{zhang2021distributed} and long baseline quantum telescopes~\cite{quantum_telescope}, but the primary use case considered here is to perform QKD at global distances~\cite{satQKD}. Importantly, when using QRs in a QKD application, the communication parties do not need to trust the intermediate nodes, as they only relay the quantum states and cannot gain any information on the distributed secure key. In this work, we investigate if a satellite-based QR link with a single intermediate node could allow the distribution of entanglement with sufficiently low noise to enable distributed quantum information processing tasks. 

To keep the presentation simple, we deliberately refrain from modeling how the transmitted and received photonic states are processed by the two communication parties Alice and Bob. For instance, for distributed quantum computing a photonic interface to a quantum computer at the ground station has to be realised, while for QKD applications it is sufficient for Alice and Bob to have access to single-photon sources (in the UL architecture) or to single-photon detectors (in the DL architecture). Here, for quantitative results we restrict ourselves to the latter case, i.e. the information at the ground stations is always stored classically.

\subsection{Downlink and uplink architecture}

In this work, we consider two architectures. The central satellite always contains two independent quantum memories (which we denote as A and B), as well as a setup to perform a Bell State Measurement (BSM) between two qubits (one in memory A and in memory B), while the support satellites either distribute entangled photon pairs or perform BSMs on incoming photon pairs. In the first architecture, the quantum communication to the ground station is performed in the downlink direction. The outer satellites are equipped with photon pair sources that distribute entangled photons to the corresponding ground station and the central satellite. See \cref{QR_Architecture}, left panel, for reference. In the second architecture, the quantum communication is in the uplink direction and the outer satellites perform BSMs on the incoming photons to perform entanglement swapping. See \cref{QR_Architecture}, right panel, for reference.

The use of memory assistance makes it possible to achieve a significant advantage in entanglement distribution rates by allowing entanglement swapping between states that have arrived asynchronously at the central repeater node. However, to achieve this a \textit{heralding mechanism} must be present: the central satellite needs information about which photons have been successfully coupled into the memory. 

Two different approaches are examined for the DL and UL architectures, which require starkly different technologies and subsystems. In the DL architecture, it is assumed that a heralding mechanism is present, i.e. either the quantum memory features an intrinsic heralding mechanism or, more generally, a Quantum Non-Demolition (QND) measurement~\cite{QND} can be performed on the central satellite to indicate the presence of a photon in the memory input without affecting the relevant properties of its quantum state.

The UL architecture, on the other hand, does not require a dedicated heralding mechanism. In this case, \textit{emissive quantum memories} are employed to create entanglement between the degrees of freedom stored in the quantum memory and those of an emitted photon. An emissive QM can be created by connecting an entangled photon pair source to a standard (absorptive) QM co-located in the central satellite (to avoid photon losses due to a long link between the source and the memory). One photon of the pair is stored in the memory while the other one is sent to one of the outer satellites where it interferes with a photon from the corresponding ground station to perform a BSM. The confirmation of a successful BSM in one of the outer satellites is used as a heralding signal to indicate that the other photon of the pair can be used for entanglement swapping at the central satellite.

\paragraph{Downlink}
The main advantage of the first architecture is that with current technology for free-space optical communication, downlinks are usually easier to realize than uplinks, i.e.\ the former have higher channel transmission~\cite{satQKD}, even when complex Laser Guide Star (LGS) systems are employed to minimize channel losses in uplink~\cite{Pugh2020}. Furthermore, in the DL architecture, the requirements on the capacity of the quantum memories (i.e., the number of modes that can be stored) are significantly lower, since the heralding signal is generated locally in the central satellite and only the photons that have generated a heralding signal need to be stored in the memory. As noted before, however, this scheme requires the use of QND measurements, which are already challenging to realize in a laboratory setting~\cite{distante2021detecting} and have never been demonstrated in space.

\paragraph{Uplink}
The main advantage of the second architecture is that it does not require a QND measurement, since the heralding mechanism is based on a BSM, which is a much more mature technology. With today's technology, a high-quality BSM can be achieved, provided that single-photon detectors with low intrinsic dark counts are employed and the arrival times of the two measured photons are precisely synchronised (so that there is a good overlap between the incoming pulses). The main challenge in signal synchronization arises as the distance between the satellites and the distance between the ground station and the outer satellites changes by several kilometers per second in LEOs. In addition, the heralding signal in the UL architecture is generated by the outer satellites and needs to be broadcast to the central satellite. Consequently, all modes have to be stored for a duration at least equal to the signal round-trip time. Therefore, a QM with a larger storage capacity is required. 

\subsection{Entanglement swapping scheme}\label{sec:swap}
To perform a BSM with a qubit pair in the two memories, the confirmation of a successful detection of the corresponding photons with which they share the entanglement must arrive at the central satellite. The time that elapses between the storage of a photon in the memory and receiving the confirmation of the arrival of its partner is referred to as the round-trip time $t_\upA^\circlearrowleft (t_\upB^\circlearrowleft)$. In the DL scenario, the ground station needs to confirm any successfully received photon to the central satellite. Due to the large distances, there is usually no direct line-of-sight between them. Therefore, the ground station sends the classical confirmation signal to the corresponding outer satellite, which relays it to the central satellite. The round-trip time is given by the difference of the photon arrivals at the ground station and the central satellite plus the time it takes the classical confirmation signal to arrive back at the central satellite. Therefore, we have
\begin{equation}
    t_{\upDL}^\circlearrowleft = \underbrace{t_\GS - t_\sat}_{\substack{\text{difference in}\\\text{photon arrivals}}} + \underbrace{t_\GS + t_\sat}_{\substack{\text{travel time of}\\\text{classical signal}}} = 2 t_\GS \,.
\end{equation}
Here, $t_\GS$ and $t_\sat$ are the travel times of the signals between the outer satellite and the ground station, and between the outer satellite and the central satellite, respectively, as indicated in \cref{QR_Architecture} for the corresponding party (Alice or Bob). In contrast, in the UL scenario, the confirmation signal is transmitted from the outer satellite to the central satellite. Here, the round-trip time is given by the time it takes the entangled photon from the central satellite to reach the outer satellite plus the time it takes the classical confirmation signal to be sent back to the central satellite, thus
\begin{equation}
    t_\upUL^\circlearrowleft = 2 t_\sat \,.   
\end{equation}
Furthermore, additional processing times may occur at each node, which we consider negligible compared to the round-trip times and do not take into account in this work.

Due to decoherence process, the quantum information stored in the quantum memories deteriorates over time. Therefore, a finite storage cutoff time $t_\upA^\upc (t_\upB^\upc)$ is introduced and only the quantum states that have been stored in memory $\upA$ ($\upB$) for a time no longer than $t_\upA^\upc (t_\upB^\upc)$ are used for the BSM in the central satellite. If multiple photons are stored within one memory before the BSM partner is registered in the other memory, the most recent one will be used for the BSM, as soon as the corresponding heralding signals arrive. In this work, we assume that the photons that had arrived before the ones employed in the BSM will be cleared from the memories.

It is assumed that the photon pair sources operate at a fixed rate $R$, with some probabilities $\eta_{\text{EPS},\upA}, \eta_{\text{EPS},\upB}$ of actually generating and emitting a pair at each trial. The entire process is then described in terms of discrete time-bins, with all events occurring at times that are (approximately) integer multiples of $1/R$. Four types of BSM events are now introduced and expanded on in appendix \ref{sec:A1}.

\paragraph{Attempted BSMs} 
If both links succeed, i.e.\ if a heralding signal is received from both memories within the respective cutoff times, a BSM is attempted between the two corresponding memory slots. For each trial, the probabilities that the individual links of Alice and Bob succeeding are denoted as $\eta_{\upA}$ and $ \eta_{\upB}$. As both links operate independently of each other, the number of trials until success is geometrically distributed for each link.

\paragraph{Successful BSMs} 
Even if both links succeeded and a BSM is attempted, the corresponding photons may have been lost during read-in, read-out or due to memory deexcitation. The deexcitation can be modeled as an exponential process, so that the probabilities of retrieving a photon from Alice's or Bob's memory after a certain time $t$ depend on the characteristic decay times of the memories $\tau_\upA, \tau_\upB$ according to
\begin{align}
    P(\checkmark|\upA, t) &= \eta_\upA^\checkmark \upe^{-t/\tau_\upA} \,, &
    P(\checkmark|\upB, t) &= \eta_\upB^\checkmark \upe^{-t/\tau_\upB} \,,
\end{align}
respectively. Here, $\eta_\upA^\checkmark$ and $\eta_\upB^\checkmark$ are the \textit{zero-time coupling efficiencies}, i.e. the probabilities that a photon has actually been coupled and can be retrieved from the memory, under the condition that a heralding signal has been received (from the QND or from the BSM on the outer satellites). A photon may also couple into the memory even if no heralding signal is received, but in absence of the heralding information these photons cannot be effectively employed in a quantum repeater scheme.

\paragraph{Correct BSMs}  
A valid BSM may either yield a correct or an erroneous result, which ultimately determines the quality of the exchange.
A flip can occur either on read-in or read-out with some characteristic probabilities $1-\eta_\upA^+$, $1-\eta_\upB^+$ or due to exponential decoherence of the memory state with some characteristic coherence times $\mathcal{T}_\upA$, $\mathcal{T}_\upB$. The probabilities of a $+$ ($-$) outcome, corresponding to a correct (erroneous) result, occurring after a certain time $t$ read
\begin{align}
    P(\pm|\upA, t) &= \frac{1 \pm \eta_\upA^+ \upe^{-t/\mathcal{T}_\upA}}{2} \,, & 
    P(\pm|\upB, t) &= \frac{1 \pm \eta_\upB^+ \upe^{-t/\mathcal{T}_\upB}}{2} \,,
\end{align}
respectively. In a partial BSM, an error occurs if exactly one of the qubits experiences a flip in their respective memory. Thus, the total probability of a correct/erroneous event after the two photons have been stored for times $t_\upA$ and $t_\upB$, ${P(\pm|t_\upA, t_\upB) = P(\pm|\upA, t_\upA) P(+|\upB, t_\upB) + P(\mp|\upA, t_\upA) P(-|\upB, t_\upB)}$, reduces to
\begin{align}
    P(\pm|t_\upA, t_\upB) &= \frac{1 \pm \eta_\upA^+ \eta_\upB^+ \upe^{-t_\upA/\mathcal{T}_\upA} \upe^{-t_\upB/\mathcal{T}_\upB}}{2} \,.
\end{align}
For simplicity, we assume that memory decoherence is the prevailing source of error and do not explicitly consider other noise contributions, such as detector dark counts or accidental multi-pair emissions from the entangled photon sources.

\paragraph{Secure BSMs}  
Additionally, as a measure of system performance that includes a trade-off between maximizing the number of successful and minimizing the number of erroneous BSMs, the number of secure BSMs
\begin{equation}
    N_\BSM^{\mathrm{sec.}} = N_\BSM^\checkmark [ 1 - H(\mathcal{E}_x) - H(\mathcal{E}_z)] = N_\BSM^\checkmark [ 1 - H(\mathcal{E}_x)]\,,
\end{equation}
is introduced in analogy to the asymptotic secure key rate in QKD applications~\cite{devetak_winter}. Here, $H$ is the binary entropy function and $\mathcal{E}_x = N_\BSM^-/N_\BSM^\checkmark$ is the ratio of erroneous to successful BSMs, corresponding to the Quantum Bit Error Rate (QBER) in the X basis, while $\mathcal{E}_z$ is the QBER in the Z basis, which we assume to be negligible ($\mathcal{E}_z=0$).

\subsection{Satellite communication link}\label{sec:comm}

\begin{figure}
    \centering
    \includegraphics{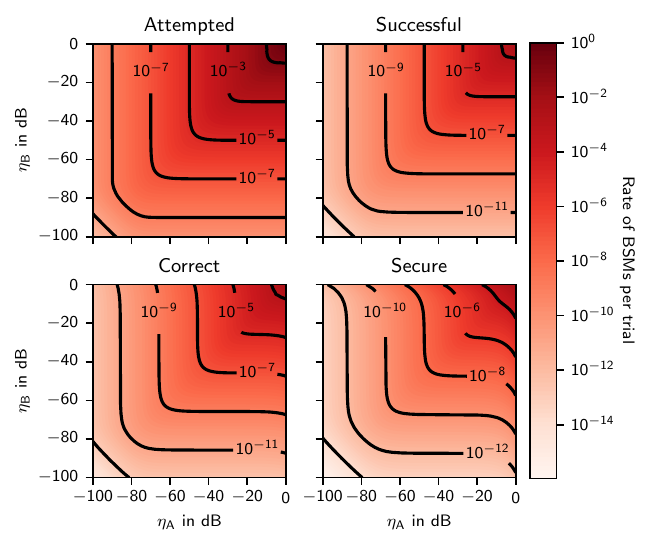}
    \caption{Rate of attempted, successful, correct and secure BSMs per trial over the total channel transmissions between Alice and the central node ($\eta_\upA$) and Bob and the central node ($\eta_\upB$). For Alice and Bob, the round-trip times are set to $t^\circlearrowleft = \SI{30}{\milli\second}$, while the characteristic decay and coherence times of the memories are $\tau = \SI{100}{\milli\second}$ and $\mathcal{T} = \SI{60}{\milli\second}$ and the memory efficiencies are $\eta^\checkmark = 0.1$. The corresponding memory cutoff times are subject to optimization.}
    \label{BSMs_transmission}
\end{figure}

One difference between satellite-based QRs compared to fiber-based QRs are the time-varying link distances. Fiber-based repeaters allow for a symmetrical design where the entanglement swapping operation occurs approximately halfway between Alice's and Bob's location.
For a given total transmission between Alice and Bob $\eta = \eta_\upA \eta_\upB$, the case of a repeater node with symmetric links maximizes the probability of performing a BSM and we have a scaling of ${P_\BSM \sim \eta_\upA = \eta_\upB}$.

For satellite-based QRs $\eta_{\upA}$ and $\eta_{\upB}$ are time-dependent, as the distances between the satellites and the ground stations change over the course of the pass. Therefore, the probability of performing a BSM at a given time scales as $P_{\BSM} \sim\text{min}(\eta_{\upA}, \eta_{\upB})$ for satellite-based repeater nodes. The worst performing repeater link $i\in \{\upA,\upB\}$ at each time step determines the overall repeater performance. For each trial, the probabilities that the individual links of Alice and Bob succeed are given by 
\begin{align} \label{eq:SingleTrialSuccessProbabilities}
    \eta_{i}^\upDL &= \eta_{\text{ch},i}^\upDL\,\eta_{\text{ch},i}^\upIS\,\eta_{\text{EPS},i}\,\eta_{\text{SPD},i}\,\eta_{\text{QND},i}, &
    \eta_{i}^\upUL &= \eta_{\text{ch},i}^\upUL\,\eta_{\text{ch},i}^\upIS\,\eta_{\text{EPS},i}\, \eta_{\text{SPD},i}^2\,\eta_{\BSM,i},
\end{align}            
for the DL and UL architecture. Here, $\eta_{\text{ch},i}$ is the channel efficiency of the satellite-to-ground (ground-to-satellite) and $\eta_{\text{ch},i}^\upIS$ the channel efficiency of the Inter-Satellite (IS) link. Furthermore, $\eta_{\text{EPS},i}$ is the photon source efficiency, $\eta_{\text{QND},i}$ the quantum non-demolition measurement efficiency, $\eta_{\text{SPD},i}$ the single-photon detection efficiency and $\eta_{\BSM,i}$ is the probability of a conclusive BSM result in the outer satellite given that two detections were registered. For a photonic BSM without ancillary photons the maximum efficiency is $\eta_{\BSM,i} = 1/2$~\cite{bsm_eff}.

\Cref{BSMs_transmission} shows the rate of attempted, successful, correct and secure BSMs over the total transmissions of Alice's and Bob's channels $\eta_\upA$ and $\eta_\upB$ for round-trip times of $t^\circlearrowleft = \SI{30}{\milli\second}$. The contour lines resemble the expected $\min(\eta_\upA, \eta_\upB)$ scaling, whilst clearly showing the impact of the limited decay and coherence times of $\tau = \SI{100}{\milli\second}$ and $\mathcal{T} = \SI{60}{\milli\second}$. For symmetric losses $\eta_\upA = \eta_\upB$, the rate is sensitive to changes in both channels. With larger asymmetries, the sensitivity of the rate to changes in the better channel decreases, while the sensitivity to changes in the worse channel increases. Precise knowledge of the link dynamics is therefore crucial for evaluating the performance of the repeater connection.

\paragraph{Channel efficiency}

\begin{figure}
\centering
\includegraphics[width=0.6\textwidth]{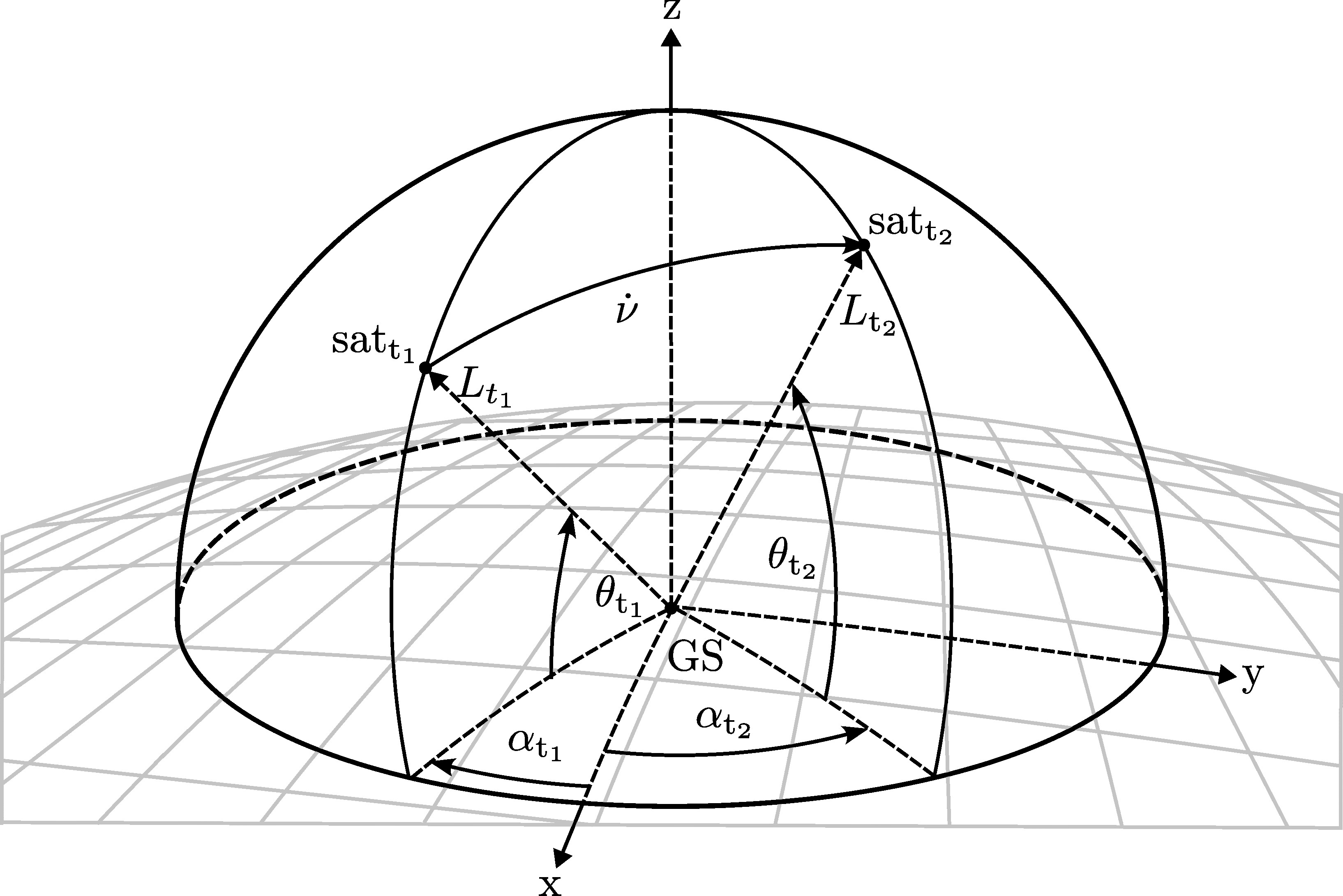}
\caption{Representation of the three-dimensional ground station pass of a single satellite at times $t_1$ and $t_2$, with elevation angle $\theta$, azimuth angle $\alpha$ and the angular velocity of the satellite relative to the ground station $\dot{\nu}$.}\label{GS_pass}
\end{figure}

The dynamics of the ground station pass which is depicted in \cref{GS_pass}, determines the channel efficiency $\eta_{\text{ch},i}$. The channel efficiency mainly depends on the link distance $L$, but also on the elevation angle $\theta$, azimuth angle $\alpha$ and, due to atmospheric turbulence effects, also on the angular velocity of the satellite relative to the ground station $\dot{\nu}$. Overflights in which $\theta$ reaches its maximum value of \SI{90}{\degree} are called zenith passes. Zenith passes lead to the highest channel efficiencies for a given satellite altitude. Depending on the link (downlink, uplink or inter-satellite link), the channel efficiency $\eta_{\text{ch},i}$ is calculated differently: 
\begin{align}
	\eta_{\text{ch},i}^\upDL & = \eta_{\text{coll},i}^\upDL\,\eta_{\text{atm},i}\,\eta_{\text{smf},i}, &
	\eta_{\text{ch},i}^\upUL & = \eta_{\text{coll},i}^\upUL\,\eta_{\text{atm},i}\,\eta_{\text{bwb},i}\,\eta_0, &
	\eta_{\text{ch},i}^\upIS & = \eta_{\text{coll},i}^\upIS\,\eta_0\label{eq:channel_ef}
\end{align} 
with the collection efficiency $\eta_{\text{coll}}$, the loss due to atmospheric attenuation $\eta_{\text{atm}}$, the single-mode fiber coupling efficiency $\eta_{\text{smf}}$, the signal loss due to beam broadening and wandering $\eta_{\text{bwb}}$ and the optical coupling efficiency $\eta_0$. Other effects that are not explicitly modeled are internal transmission losses due to finite reflectivity and transmissivity of mirrors and other optical elements within the terminals, as well as pointing losses due to platform vibrations.

\paragraph{Collection efficiency}
The collection efficiency of an optical receiver (e.g., a telescope) is given by
\begin{align}
    \eta_{\text{coll}} = \frac{1}{P_\Tx}\int_{\mathcal{A}_\Rx} I(x,y) \upd x \upd y,
\end{align}
where $I(x,y)$ is the intensity of the electromagnetic radiation at the position $(x,y)$ on the receiver plane, $\mathcal{A}_\Rx$ is the receiver aperture and $P_\Tx = \int_{\mathds{R}^2}  I(x,y) \mathrm{d}x\mathrm{d}y$ is the total transmitted power. If the receiver is placed in the far field of the transmitter (i.e., the spot size at the receiver is much larger than the receiver aperture) and aligned with the axis of the beam, the integral converges to $\int_{\mathcal{A}_\Rx}  I(x,y) \mathrm{d}x\mathrm{d}y = I_\text{peak} A_\Rx$, where $I_\text{peak}$ is the peak intensity of the beam and $A_\Rx$ is the area of the receiver aperture. Then the collection efficiency in the far field is
\begin{align}
	\eta_{\text{coll,ff}} = 
	\frac{I_\text{peak} A_\Rx}{P_\Tx} = 
	\overbrace{\frac{I_\text{peak}}{\frac{P_\Tx}{4\uppi L^2}}}^{G_\Tx}
	\overbrace{\frac{\lambda^2}{\left(4\uppi L\right)^2}}^{\eta_{\text{fs}}}
	\overbrace{\frac{4\uppi A_\Rx}{\lambda^2}}^{G_\Rx} = 
	G_{\Tx} G_{\Rx} \eta_{\text{fs}} 
	\label{eq:eta_col}.
\end{align}
Here, $\lambda$ is the wavelength and $L$ is the link distance~\cite{Degnan1974}. Furthermore, we have introduced the transmitter gain $G_{\Tx}$, the receiver gain $G_{\Rx}$ and the free-space loss $\eta_\text{fs}$ or geometric loss. The transmitter gain $G_\Tx$ corresponds to the ratio between the peak intensity of the radiation pattern and the intensity that an isotropic radiator would have. From Fourier optics we have ${G_\Tx} = {I_\Tx/I_\text{iso}} \leq {4\uppi A_\Tx/\lambda^2}$. If the receiver is not sufficiently smaller than the spot size, a correction factor for the collection efficiency can be derived, yielding
\begin{align}
	\eta_{\text{coll}} =
	1-\upe^{-\eta_{\text{coll,ff}}},
\end{align}
as presented in appendix \ref{sec:coll}.

\paragraph{Atmospheric losses in the downlink and uplink channel}

Most of the atmospheric attenuation occurs in the \SI{20}{\kilo\meter} closest to the surface of Earth, where the atmosphere is densest, and affects the UL and DL signals equally. Atmospheric attenuation is strongly weather-dependent, with the presence of clouds resulting in complete blocking of the signal, but for simplicity we consider only a nominal case with $\SI{23}{\kilo\meter}$ of horizontal visibility. The signal attenuation increases at lower elevation angles as the signal has to travel longer distances through the atmosphere. That is, if the link has an elevation angle $\theta\geq\SI{20}{\degree}$ above the horizon, the atmospheric transmission can be well approximated as~\cite{hemmati2020near}
\begin{align}
    \eta_{\text{atm}}(\theta) = \eta_{\text{atm,zen}}^{1/\sin(\theta)},
\end{align}
where $\eta_{\text{atm,zen}}$ is the atmospheric transmission from ground to space in the zenith direction. A model that holds also for lower elevation angles can be found in~\cite{Kasten:89}.

\paragraph{Single-mode fiber coupling}
After the photons are received by an optical terminal they have to be coupled into a Single-Mode Fiber (SMF) to be expediently routed to the quantum information processing subsystems, such as the quantum memories and BSM devices. The tail of the optical fiber is placed in the focal plane of the optical terminal, centered on the spot where the received light is focused on. The coupling efficiency is given by the overlap between the electromagnetic field of the free-space signal and the one supported by the SMF.

Wavefront distortions are the main cause of SMF coupling inefficiency. Atmospheric turbulence creates air pockets with different temperatures, leading to variations in refractive index, which then result in wavefront distortions when the beam propagates through the lower part of the atmosphere. The characteristic length of the wavefront distortions is expressed by the spatial coherence radius, also known as Fried parameter~\cite{fried1966optical}. In the DL channel, the Fried parameter is typically in the order of a few centimeters. If the Fried parameter is small compared to the receiver diameter this leads to a starkly reduced coupling efficiency and Adaptive Optics (AO) systems are needed to achieve a good SMF coupling, see appendix \ref{sec:SMF} for a short overview. As the Fried parameter depends on the elevation angle, the fiber coupling efficiency changes during the satellite pass and takes the lowest value at small elevation angles~\cite{Scriminich2022}.

For an UL signal to LEO satellites the beam expands while traveling for several hundred kilometers after exiting the denser part of the atmosphere. The expansion of the beam also leads to a corresponding expansion and smoothing of the wavefront distortions~\cite{Andrews2005} and by the time the signal reaches the satellite, the wavefront distortions are several meters in size. The effects of the distortions on the coupling loss are negligible and the coupling inefficiency is only given by the mismatch between the point spread function of the receiver telescope and fiber field shape. The point spread function of a circular aperture is an Airy disk, while the SMF transmitted mode is Gaussian, resulting in a maximum optical coupling efficiency of around $\eta_0 = 81\%$~\cite{toyoshima2006}. However, by using advanced beam forming technologies within the receiver terminal almost unit SMF coupling efficiency can be achieved~\cite{ellis2021, bouhafs2022}.

\paragraph{Beam wandering and beam broadening}
For the UL channel, small turbulence cells, which are narrower than the beam transversal size, refract different portions of the beam in different directions and thus also result in an increase of the (instantaneous) beam divergence. Larger atmospheric turbulence cells instead cause a coherent refraction of the beam. The beam is diverted from its original path with a deviation that changes over time on a millisecond time scale, resulting in so-called beam wandering. The power received at the satellite fluctuates over time (depending on how much the beam is refracted) and is reduced in mean intensity, which can be interpreted as an increase in the (long-term averaged) beam divergence. 

Conventional methods to counteract these effects, such as pointing-by-tracking beam stabilization, are currently being investigated for LEO satellites~\cite{Piazzolla2023}, where the point-ahead angle reaches up to \SI{50}{\micro\radian}. The challenge is that for fast-moving satellites, the required point-ahead angle is often larger than the so-called isoplanatic angle -- the angular region over which the turbulence effect do not change appreciably. This means that the UL and DL beams encounter different turbulence conditions, rendering the channels non-reciprocal. Essentially, while the downlink beam is exposed to certain atmospheric conditions, the uplink beam could encounter entirely different turbulence patterns. This leads to less correlation between the uplink signal and the beacon used to correct the wavefront distortions. 

To improve the performance in such scenarios, an LGS system can be implemented that enables the reception of a reference signal that follows the same path through the atmosphere as the uplink signal. We model the achievable coupling efficiency according to the references~\cite{Tyson2015, Pugh2020}, see appendix \ref{sec:bwb}.

\paragraph{Total channel efficiency and elevation angle}\label{sec:tot-eff}
By modeling the effects outlined in the previous paragraphs, it is possible to determine the total channel efficiency defined by \cref{eq:SingleTrialSuccessProbabilities} for each link. For inter-satellite links, the only dynamical parameter affecting the channel efficiency is the link distance $L$. The channel efficiency decreases quadratically with the distance due to beam divergence, see \cref{eq:eta_col}. For UL and DL the channel efficiency is maximum when the satellite is at the zenith of the ground station to which it is connected. When the elevation angle of the satellite decreases the channel efficiency gets worse. This is mainly caused by the longer link distance, but also due to increased atmospheric absorption and stronger turbulence effects. At very low elevations, it is therefore difficult to establish a link of sufficient quality. We take this into account in our simulation by discarding the optical links below a certain minimum elevation angle $\theta_{\min}$.

\section{Satellite dynamics}\label{sec:sat}

\begin{figure}
\centering
\includegraphics[width=1.0\textwidth]{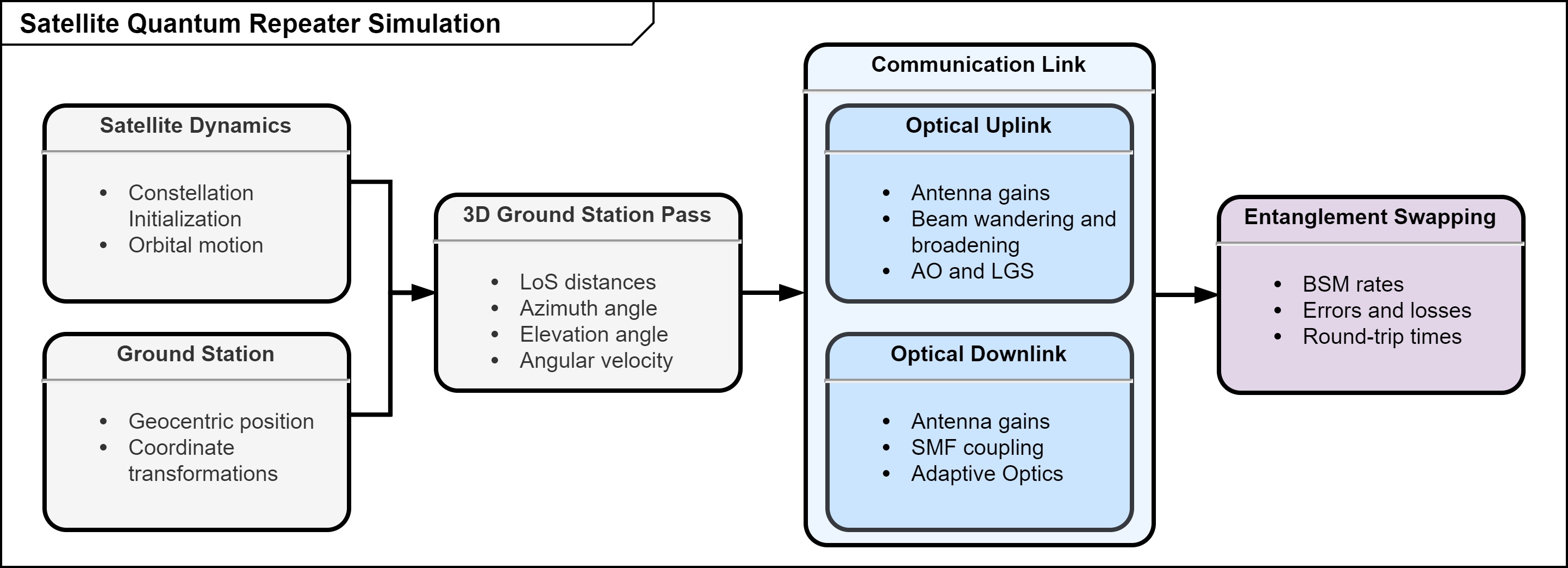}
\caption{Illustration of the high-level software architecture. Based on the equations of motion, future satellite positions are calculated from their initial conditions. Together with the locations of Alice's and Bob's ground stations the geometry of the three-dimensional ground station pass can be computed, which determines the channel loss of the communication link. Depending on the repeater architecture, either a downlink or an uplink channel is established. In the last step, the entanglement swapping scheme is carried out, which provides the final BSM rates.}\label{Sim_Scheme}
\end{figure}

\Cref{Sim_Scheme} shows the high-level software architecture of our integrated simulator, which combines satellite orbit propagation with communication channel efficiency calculations and the entanglement swapping operation. Using this model, we can initialize the satellite constellations and choose arbitrary sites for Alice's and Bob's ground stations. We take into account the rotation of Earth and effects of its non-spherical gravitational potential on the orbit of the satellite. Depending on the position of the satellite and the ground station coordinates, the geometric parameters of the ground station passes are calculated. These parameters are used to evaluate the downlink and uplink channel efficiency, including adaptive optics corrections, see \cref{sec:comm} and appendix \ref{sec:A2} for more information. With the channel efficiencies, round-trip times and the satellite state vector, the BSM rate at each satellite position can be determined, see \cref{sec:swap} and appendix \ref{sec:A1}. Furthermore, by calculating the position of the Sun with respect to Earth we can distinguish between nighttime and daytime passes. This allows us to specify additional requirements e.g., that only ground station passes during local nighttime should be considered for the repeater performance.

\subsection{Orbit initialization}

\begin{figure}
\centering
\includegraphics[width=0.6\textwidth]{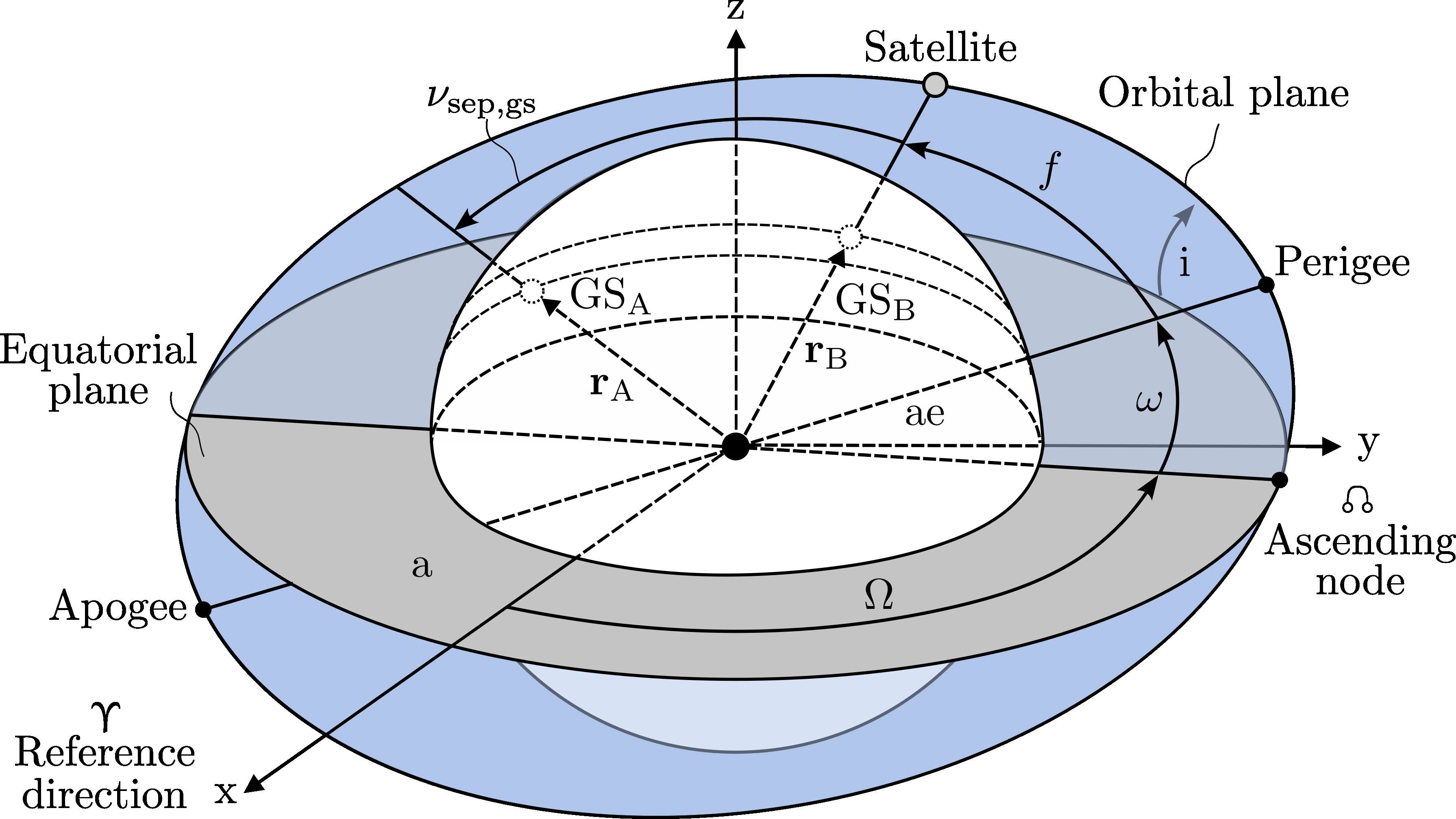}
\caption{Sketch of an Earth-centered coordinate system expressed in Kepler orbit elements, including the locations of Alice's and Bob's ground stations. The equatorial plane is colored grey and the orbital plane blue. The x-axis points to a celestial reference point.}\label{Orbit_Geo}
\end{figure}

Using satellite dynamics simulations, we can analyze arbitrary orbits. To define the orbit and initial positions of the satellites in the constellation, we introduce Kepler orbit elements, see \cref{Orbit_Geo}. Kepler orbit elements describe the orbit geometry and the position of the spacecraft using six independent variables. The \textit{eccentricity} $e$ and the \textit{semi-major axis} $a$ define the shape and size of the orbit. The \textit{inclination} $i$, \textit{Right Ascension of the Ascending Node} (RAAN) $\Omega$ and the \textit{argument of perigee} $\omega$ characterize the rotation of the orbit in space. The \textit{true anomaly} $f$ indicates the position of each satellite on the orbit at a certain epoch~\cite{Steiner2004}. Furthermore, the position of the ground stations is defined by \textit{latitude} ($\upphi$) and \textit{longitude} ($\uplambda$).

To conveniently describe the relative position of satellites in a constellation, we introduce the additional parameter $\nu_{\text{sep,sat}}$ expressing the angular separation between the outer satellites as measured at the center of Earth. If the angular separation is identical to the angular separation of the two ground stations ($\nu_{\text{sep,gs}}=\nu_{\text{sep,sat}}$) the outer satellites can be aligned at the zenith of both ground stations, see \cref{Orbit_Geo}. Therefore, the separation between the outer satellites $\nu_{\text{sep,sat}}$ is expressed relative to the angular separation of the ground stations $\nu_{\text{sep,gs}}$. The central satellite is positioned in the middle between the outer satellites and is separated by approximately $\nu_{\text{sep,sat}}/2$ from both of them.

\subsection{Orbit propagation} \label{sec:dynamics}

By propagating the orbit of each satellite based on the initial state vector, future satellite positions are predicted using the equations of motion of a two-body system. Forces that do not originate from the gravitational acceleration of Earth but influence the satellite dynamics are called disturbance forces. In the case of a LEO satellite, these include non-gravitational forces such as atmospheric drag and solar radiation pressure~\cite{Steiner2004}. Compensating for these disturbances is the task of the satellite Attitude and Orbit Control System (AOCS). In this work, we assume that the AOCS is able to perfectly compensate for the disturbance forces, hence we only consider the gravitational acceleration of Earth. 

The gravitational potential of Earth can be treated as axisymmetric and can be expanded in terms of zonal harmonic coefficients $J_n$. These constants carry the information about the mass distribution of a rotationally symmetric body and must be determined experimentally. The gravitational potential is~\cite{Steiner2004}
\begin{equation}
    V(r,\upphi) = -\frac{GM_{\text{E}}}{r}\bigg[1-\sum^{\infty}_{n=2}\bigg(\frac{R_{\text{E}}}{r}\bigg)^nJ_nP_{n}(\sin{\upphi})\bigg],
\end{equation}
with the satellite distance $r$ to Earth's center of mass, the latitude $\upphi$, the gravitational constant $G$, the mass $M_{\text{E}}$ and the equatorial radius of Earth $R_{\text{E}}$, and the Legendre polynomials $P_{n}$~\cite{Steiner2004}. The inclusion of the coefficient $J_2$ is sufficient to capture the most important effects, such as the precession of the orbit due to the flattening of Earth's poles. It is three orders of magnitude larger than the other coefficients and primarily effects the Kepler elements RAAN $\Omega$, argument of perigee $\omega$ and true anomaly of the satellite $f$~\cite{fumenti2020}. While the $J_2$ coefficient has little influence on a single satellite pass, it is of crucial importance for long-term simulations of the constellation dynamics.

\section{Performance analysis}\label{sec:performance}

We define two operating scenarios for the performance analysis of the QR. The first scenario describes an intercontinental (IC) link between a ground station in New York City and Berlin. The ground stations are \SI{6385}{\kilo\meter} apart and separated by $\nu_{\text{sep,gs}} =  \SI{57.6}{\degree}$. The second scenario simulates a link between two cities in Europe (EU), Madrid and Berlin. This distance is \SI{1870}{\kilo\meter} with an angular separation of $\nu_{\text{sep,gs}} =  \SI{16.8}{\degree}$. For the following analyses, we consider a minimum elevation angle of $\theta_{\text{min}} = 20^\circ$. The communication wavelength is $\lambda=\SI{1550}{\nano\meter}$, which exhibits a good atmospheric transmission. The ground station telescopes have a diameter of $D_{\text{GS}} = \SI{1}{\meter}$ and the satellite telescopes have a diameter of  $D_{\text{Sat}} = \SI{0.5}{\meter}$. A detailed overview of the communication link and QR parameters can be found in appendix \ref{sec:Param}.

We start with the investigation of single zenith passes, which can be seen as the ideal case of a three-dimensional ground station pass. Afterwards, we look at the effect of different satellite altitudes and inter-satellite distances on the BSM rate. Finally, we determine the annual repeater performance for two satellite orbit configurations.  

\subsection{Zenith passes}

In the first analysis, we only consider the results for the IC scenario, as both scenarios exhibit similar behaviour with regard to zenith passes. The initial positions of the satellites are chosen so that both outer satellites reach the zenith $\theta=90^\circ$ above their respective ground station in New York City and Berlin at the same time. The satellites are placed on circular LEOs at an altitude of \SI{500}{\kilo\meter} so that they follow each other with an angular separation of $\nu_{\text{sep,sat}}=\SI{28.8}{\degree}$. 

\paragraph{Channel loss}

\begin{figure}
\centering
\includegraphics{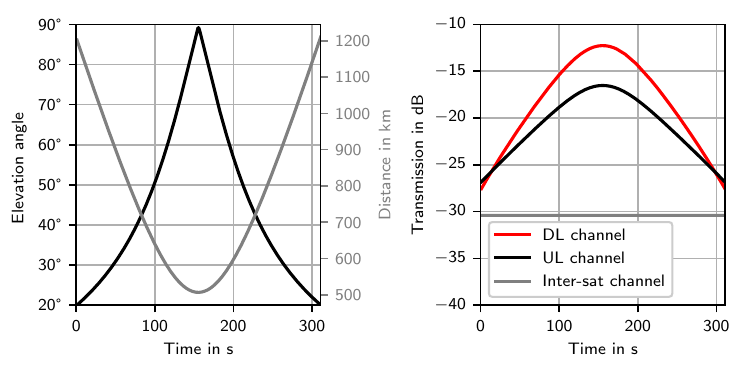}
\caption{Results of the single zenith pass at Alice's ground station for the DL and UL architecture at \SI{500}{\kilo\meter}. (Left) Link distance $L$ and elevation angle $\theta$ as a function of time. The maximum elevation angle is reached when the satellite is directly above the ground station. (Right) Transmission of the DL, UL and inter-satellite channels as a function of time. DL (UL) transmission varies greatly over the course of the zenith pass, as it depends on the link distance and elevation angle. The inter-satellite transmission is constant, as it only depends on the invariant inter-satellite distance.}\label{TransmissionLoss_Distance_IC}
\end{figure}

The left panel of \cref{TransmissionLoss_Distance_IC} depicts the elevation angle and link distance of the satellite connected to Alice's ground station. The ground station pass takes a total of \SI{311}{\second}. The link is established at an elevation angle of \SI{20}{\degree}, with a maximal link distance of \SI{1200}{\kilo\meter}. The elevation angle increases until the satellite reaches the zenith at $\theta = \SI{90}{\degree}$, where the link distance approaches the satellite altitude of \SI{500}{\kilo\meter}.

The right panel of \cref{TransmissionLoss_Distance_IC} shows the communication loss associated with Alice's ground station pass. The inter-satellite communication loss of \SI{30.4}{dB} is constant during the ground station pass, because it only depends on the static inter-satellite distances. On the other hand, the DL and UL losses vary greatly due to significant changes in the distance to the ground station and Line-of-Sight (LoS) angle. When the satellite initially connects to the ground station, the communication distance is maximum and the elevation angle is minimum, resulting in losses of about \SI{27.6}{dB} and \SI{26.8}{dB} for the DL and UL channel. Low angles mean long signal paths through the atmosphere, leading to a higher number of accumulated wavefront errors for both the DL and UL architectures. Minimum losses are observed at the zenith, with \SI{12.3}{dB} and \SI{16.5}{dB}, for the DL and UL loss. The total channel loss is then the sum of the losses in the DL (UL) channel and inter-satellite channel.

\paragraph{Bell state measurement rate}

The BSM rate depends on the time-varying link efficiency and link distance. Given the repetition rate of $\SI{90}{\mega\hertz}$ the attempted, successful, correct and secure BSM rates are calculated. It is assumed that both on Alice's and Bob's side the system works with the same QMs, entangled photon sources and single-photon detectors, therefore the corresponding indices are omitted.

\begin{figure}
\centering
\includegraphics{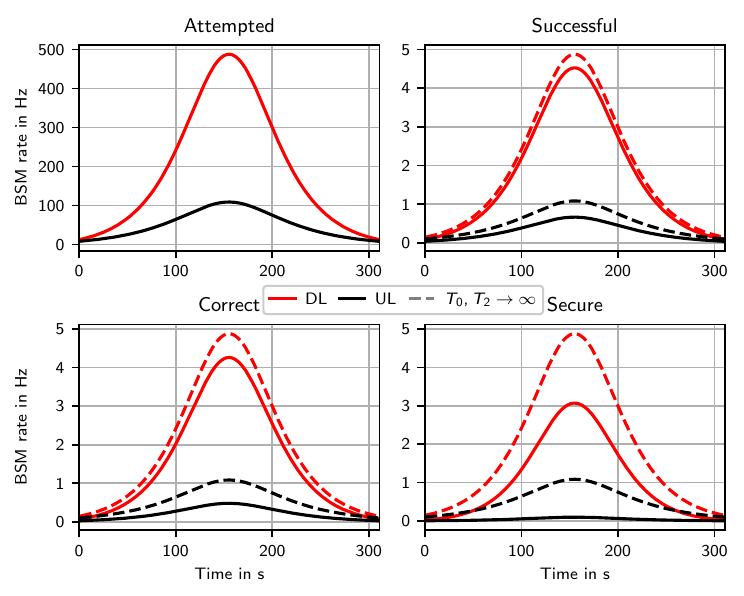}
\caption{Comparison of BSM rates between the DL and UL scenarios for a single zenith pass at a satellite altitude of \SI{500}{\kilo\meter}. (Top left) Number of attempted BSMs per second over time. The maximum rate is reached at zenith. (Top right) Number of successful BSMs per second over time. The dashed curves show the successful BSMs when the characteristic decay time of the memory is set to infinity. (Bottom left) Number of correct BSMs per second over time. The dashed curves depict the resulting BSM rate when characteristic decay and coherence times are set to infinity. (Bottom right) Number of secure BSMs per second over time. The secure BSM rate is used as a performance indicator and corresponds to the amount of correct and secret bits.}\label{BSM_AttSuccCorrSec}
\end{figure}

The upper left graph of \cref{BSM_AttSuccCorrSec} shows the attempted BSMs, i.e. whether a link is successful within the cutoff times $t_\upA^\upc, t_\upB^\upc$. The attempted BSM rate is close to zero at $t = \SI{0}{s}$ when the link is established. As the satellites move towards their respective ground stations, the rate increases up to \SI{487.8}{Hz} and \SI{108.6}{Hz} for the DL and UL architecture. The lower rate in attempted BSMs for the UL architecture is related to the higher channel loss and the lower efficiencies of the contributions to the individual link success probabilities, see~\cref{eq:SingleTrialSuccessProbabilities}.

The losses caused by the imperfect coupling to the QM determine whether a photon can be written into the memory and read out again later. The resulting reduction in success probability is given by the efficiency of the memory. With a memory efficiency of $\eta^\checkmark = 0.1$ for both memories and an infinite decay time $\tau$, the expected rate drops by a factor of $(\eta^{\checkmark})^2=0.01$. This leads to a maximum successful BSM rate for the DL and UL architecture of \SI{4.88}{Hz} and \SI{1.09}{Hz}, depicted by the dashed curves. Introducing a finite characteristic decay time $\tau = \SI{100}{\milli\second}$ results in exponential photon loss in Alice's and Bob's memories, leading to an additional decrease in the successful BSM rate. The solid lines in the upper right graph show the successful BSM rates for a finite $\tau$, with the DL architecture reaching up to \SI{4.53}{\hertz} and the UL architecture \SI{0.67}{\hertz}. The exponential loss is determined by the round-trip time and differs for the DL and UL architecture. The round-trip time of the UL architecture is \SI{22.8}{\milli\second} as it is defined by the inter-satellite distance. The round-trip times in the DL architecture depend on the link distances to the ground stations, which are much shorter and therefore result in value of \SIrange[]{3.4}{8.1}{\milli\second}. 

Due to exponential decoherence and flips during read-in and read-out, a successful BSM can be erroneous. In order to take the erroneous measurements into account, the characteristic coherence time of the memory $\mathcal{T} = \SI{60}{\milli\second}$ is included in the calculation of the correct BSMs. Here too, a longer round-trip time leads to higher number of errors, reducing the rate to \SI{4.26}{\hertz} and \SI{0.48}{\hertz} for the DL and UL architecture, as shown by the solid curves in the lower left plot. The rate of secure BSMs corresponds to the amount of correct and secret bits that could be extracted when using the entangled pairs for QKD in the asymptotic limit. The high number of errors in combination with the low number of successful BSMs leads to an even larger difference between both architectures. For the DL, we obtain a maximum rate of secure BSMs of \SI{3.07}{Hz}, while for the UL we receive \SI{0.1}{Hz}. 

The results show that for these parameters, the final rate difference between the DL and UL architecture is determined not only by the lower link efficiency of the uplink channel but also by the longer round-trip times. Choosing an architecture that minimizes the round-trip time can reduce the requirement on the storage time of the QM, especially for long communication distances. In the following sections, we will therefore continue our analysis with the DL architecture.

\subsection{Impact of satellite altitude and inter-satellite distance}\label{Sat_distance_alt}

In this section, we analyze the effect of the satellite altitude $h$ and the inter-satellite distance $\nu_{\text{sep,sat}}$ on the DL repeater performance of a single ground station pass for the IC and EU scenarios. Again, the satellites follow each other on circular LEOs, however the satellite altitudes are varied between \SI{500}{\kilo\meter} and \SI{1000}{\kilo\meter}. In addition, the inter-satellite distance is varied between $\nu_{\text{sep,sat}}/\nu_{\text{sep,gs}} = 1$, meaning the outer satellites are separated by the same angle as the two ground stations, and $\nu_{\text{sep,sat}}/\nu_{\text{sep,gs}} = 0$, meaning the outer satellite are placed at the location of the central satellite.

\begin{figure}
\centering
\includegraphics{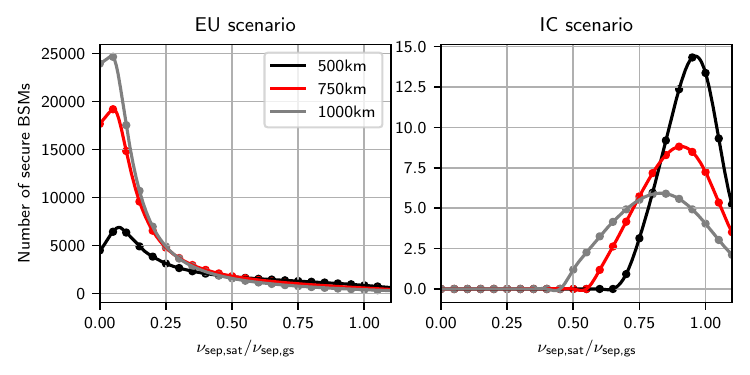}
\caption{Analysis of the number of BSMs during a single pass for the EU/IC downlink scenario as a function of the inter-satellite distance. (Left) The EU scenario demonstrates that reducing the inter-satellite distance leads to higher numbers of secure BSMs. (Right) In contrast, the IC scenario shows higher numbers of secure BSMs for inter-satellite distances closer to $\nu_{\text{sep,sat}} = \nu_{\text{sep,gs}}$.}\label{BSM_distance_down}
\end{figure}

\Cref{BSM_distance_down} shows the number of secure BSMs accumulated during a single ground station pass as a function of the inter-satellite distance for different satellite altitudes. As the satellite altitude increases from \SI{500}{\kilo\meter} to \SI{1000}{\kilo\meter}, the number of BSMs rises as higher elevation angles are reached and the link duration increases from \SI{62}{\second} to \SI{295}{\second} for $\nu_{\text{sep,sat}}/\nu_{\text{sep,gs}} = 0.1$. For the EU scenario, the best performance is achieved when the outer satellites are moved towards $\nu_{\text{sep,sat}}/\nu_{\text{sep,gs}} = 0.05$ at an altitude of \SI{1000}{\kilo\meter}, resulting in 22610 secure BSMs for the entire zenith pass. At $\nu_{\text{sep,sat}}/\nu_{\text{sep,gs}} = 0.1$ we approach the near-field regime, in which the inter-satellite channel efficiency only increases slightly, see appendix~\ref{sec:coll}. After the number of BSMs peaks at $\nu_{\text{sep,sat}}/\nu_{\text{sep,gs}} = 0.05$, moving the satellites to the center leads to a further reduction in connection time and a decrease in the BSM rate. Furthermore, in this case entanglement distribution from a single satellite is also possible, which would outperform a three-satellite quantum repeater as no quantum memories are required.

In this scenario, the communication distances are short enough to allow for direct LoS to the ground stations. The findings indicate that it would be more advantageous to send the signal directly to a single satellite in the center instead of using a constellation with three satellites.

The longer link distance of the IC scenario leads to lower channel efficiencies in the inter-satellite link and long photon storage times, which lowers the number of BSMs compared to the EU scenario. In addition, the long communication distances do not allow a direct LoS to the outer satellites when they are moved towards the center. At $\nu_{\text{sep,sat}}/\nu_{\text{sep,gs}} = 0.5$, direct LoS contact is not possible for any of the investigated altitudes and the BSM rate drops to zero. 

In contrast to the EU scenario, the IC scenario has higher rates when the constellation is in a lower orbit with a inter-satellite distance close to $\nu_{\text{sep,sat}}/\nu_{\text{sep,gs}} = 1$, as it minimizes the link distance between ground stations and satellites. Here, the maximum number of secure BSMs is $344$ for the entire zenith pass. The number of secure BSMs reduces sharply when $\nu_{\text{sep,sat}}/\nu_{\text{sep,gs}}$ is decreased beyond $\nu_{\text{sep,sat}}/\nu_{\text{sep,gs}} = 0.95$. Furthermore, there is no advantage in increasing the distance between the satellites above $\nu_{\text{sep,sat}}/\nu_{\text{sep,gs}} = 1$.

The analysis shows that the advantage of a repeater constellation can be leveraged particularly for the IC scenario with a ground station distances of \SI{6385}{\kilo\meter}, while for the EU scenario with a ground station distances \SI{1870}{\kilo\meter} a single satellite on a higher LEO could outperform the repeater constellation. Therefore, we concentrate our final analysis on the IC scenario with a DL architecture.

\subsection{Annual rate of Bell state measurements} \label{sec:year}

\begin{figure}
\centering
\includegraphics[width=1\textwidth]{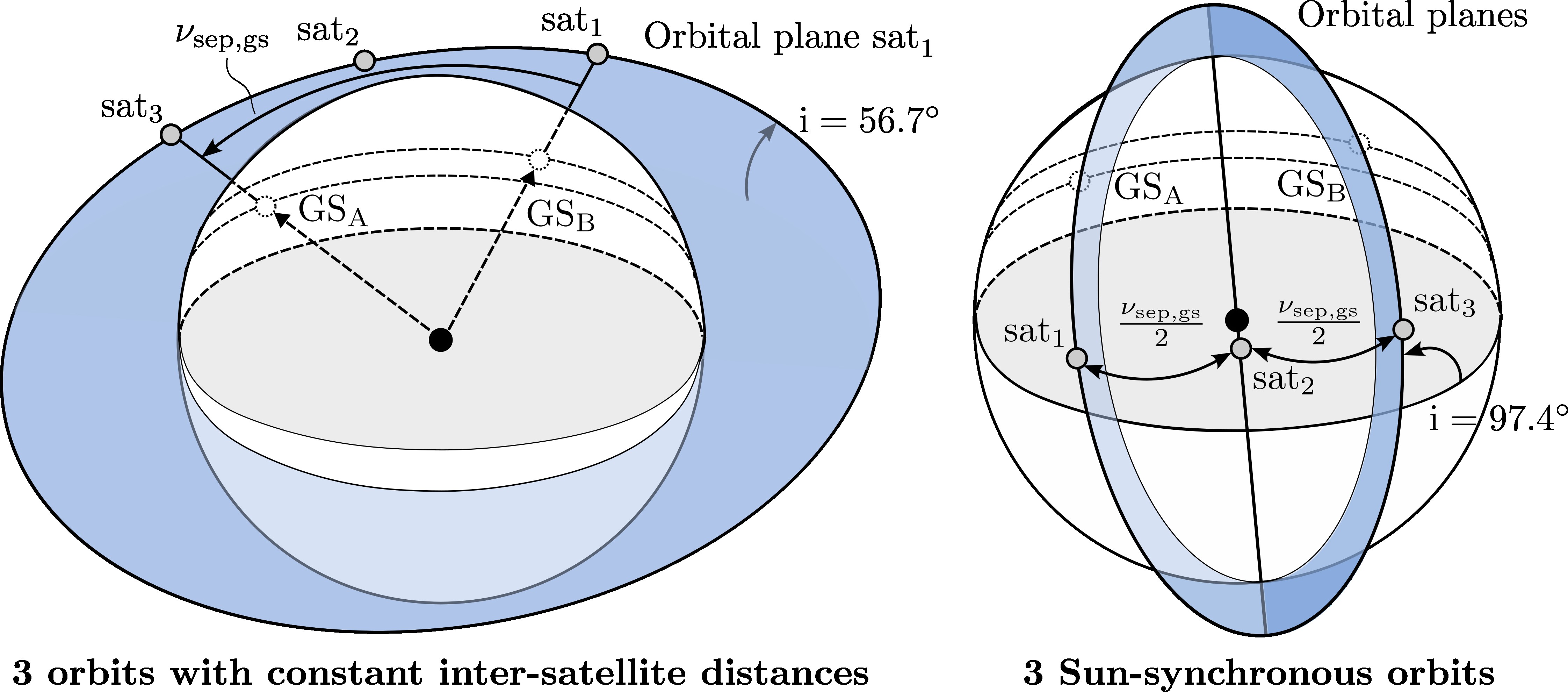}
\caption{Illustration of the two orbital configurations for the QR constellation. (Left) The figure shows the orbit configuration with constant inter-satellite distances. Only the orbit of the first satellite is depicted, as the differences to the other two orbits are small. (Right) The figure presents the orbit configuration using satellites on three Sun-synchronous orbits.}\label{Orbits_constellation}
\end{figure}

Due to strongly varying link dynamics for different ground station passes, it is necessary to analyze long time periods, usually one year, to evaluate the performance of the satellite repeater. While we do not consider the effect of cloud coverage at the ground station locations, we include the effects of the rotation of Earth and the J2 perturbation.

As described in \cref{sec:dynamics} the $J_2$ coefficient affects the orbital elements $\Omega$, $\omega$ and $f$, which can cause the satellites in the repeater constellation to drift apart, leading to a loss of connection. Therefore, one possibility is to choose the initial conditions of the satellites in such a way that the inter-satellite distance is not or only slightly influenced by the J2 perturbation. \Cref{Orbits_constellation} left panel shows this particular orbit configuration. For ground stations in New York City and Berlin, each of the three orbits must be inclined by approximately \SI{56.7}{\degree} to ensure a possible alignment. All the satellites are equidistant from each other in the along-track direction with $\nu_{\text{sep,sat}} = \nu_{\text{sep,gs}} = \SI{57.6}{\degree}$. 

\Cref{Orbits_constellation} right panel shows the second orbit configuration. Orbits where the RAAN changes at the same rate as Earth orbits around the Sun are called Sun-Synchronous Orbits (SSO). These orbits require specific inclinations for each satellite altitude, but offer the advantage that the satellite passes the ground station at approximately the same local mean solar time. Appendix \ref{sec:Param} gives the initial conditions for the satellite orbits.

\begin{figure}
\centering
\includegraphics[width=1\textwidth]{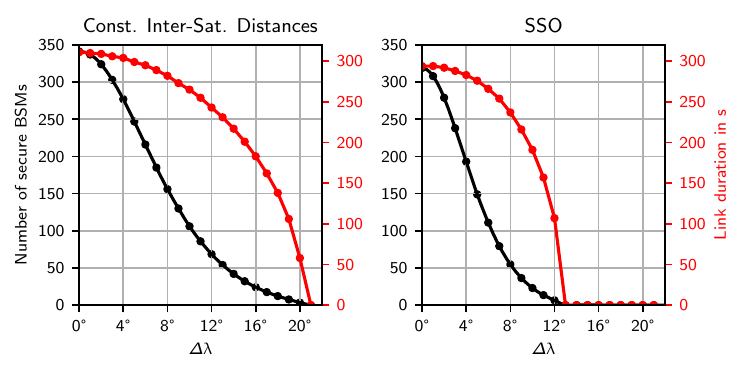}
\caption{Analysis of the connection time and the total number of secure BSMs for different angles around the rotational axis of Earth, with the zenith pass at $\Delta\lambda_\upA=0^\circ$. It shows that the number of secure BSMs reduces sharply if the satellite does not pass the zenith of the ground station. (Left) Orbit configuration with constant inter-satellite distance. (Right) Orbital configuration with three SSOs.}\label{Erot}
\end{figure}

Using these two orbit configurations, we now analyze the effect of the rotation of Earth on the link performance, which changes the geometric parameters of each consecutive ground station pass, resulting in fluctuating secure BSM rates. \Cref{Erot} shows the number of BSMs of a single pass for different shifts in longitude $\Delta\uplambda_\upA$ compared to the case of a zenith pass. The longitude difference between the ground track position of the satellite and Alice's ground station changes at a rate of about $\SI{15}{\degree/\hour}$ as the Earth rotates.

We see similar behavior for both orbit configurations. The number of secure BSMs (black curves) decrease rapidly with increasing $\Delta\uplambda_\upA$. If the deviation from the zenith pass is $\Delta\uplambda_\upA = 5^\circ$, the total number of BSMs drops by 29\% and 54\% for the orbit configuration with constant inter-satellite distances and SSOs, respectively. At the same time, the link duration (red curves) remains at a high level and only reduces by 3.5\% and 6.5\%. Thus, the main reason for the reduction in secure BSMs is the longer average link distance, resulting in higher channel losses. Furthermore, the distance between the ground track and the ground station increases faster in the constellation configured with SSOs, making it more sensitive to $\Delta \uplambda_\upA$ deviations from the zenith pass. This shows that LEO orbits with non-zenith passes must also be considered when evaluating the performance of a satellite-based quantum repeater.

\begin{figure}
\centering
\includegraphics{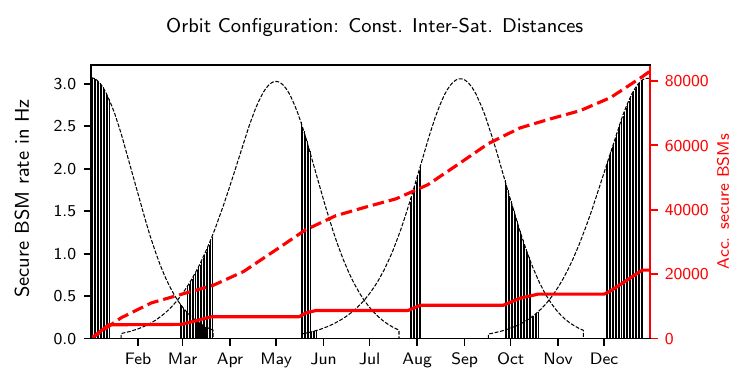}
\caption{Analysis of the annual performance of the orbit configuration with constant inter-satellite distance at an altitude of \SI{500}{\kilo\meter}. The ground stations are located in New York City and Berlin. The black vertical lines depict the the maximum secure BSM rate for each ground station pass, demonstrating its variations over the course of a year. As the ground station passes must take place at night, there are gaps in the graph. The dashed black curves represent the envelope of all possible connections, including daytime and nighttime passes. The red solid curve shows the accumulated secure BSMs when only nighttime passes are considered and the red dashed curve gives the combined number for daytime and nighttime passes. A total of 82968 secure BSMs are performed at the central satellite, of which 21268 take place during the night.}\label{BSM_DL_NB_inc}
\end{figure}

\Cref{BSM_DL_NB_inc} shows the repeater performance of the orbit configuration with constant inter-satellite distances for a time period of one year. The black vertical lines represent the maximum secure BSM rate of each ground station pass and the red curve gives the accumulated number of secure BSMs. The initial conditions of the satellite constellation are optimized for January 1st and therefore lead to the highest rates in winter. The maximum secure BSM rate is \SI{3.07}{\hertz} analogous to the result of a single zenith pass, demonstrating that the conditions for a zenith pass are only met a few times a year. In total 21268 secure BSMs are achieved. 

We can see that if only the nighttime passes are taken into account, there appear several week-long connection gaps over the course of a year. The reason for this is that the J2 perturbation rotates the ascending node $\Omega$ around the center of Earth. If the rotation of the ascending node is not synchronized with Earth orbiting around the Sun, there are days when no ground station passes occur during nighttime, because at least one ground station is in daylight. If the system would also be capable of operating in daylight, connections throughout the entire year are possible, represented by the black dashed curve. The red dashed curve gives the total number of secure BSMs of 82968. Alternatively, an additional repeater constellations with initial conditions that are optimized for the summer months, could be instaled to shorten the connection gaps.

\begin{figure}
\centering
\includegraphics{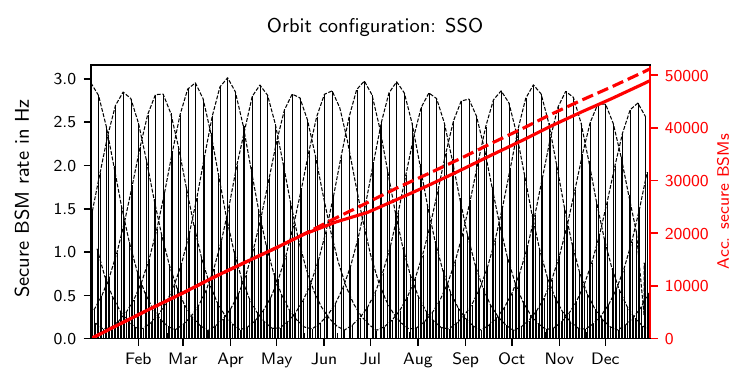}
\caption{Analysis of the annual performance of the orbital configuration with SSOs at an altitude of \SI{500}{\kilo\meter}. The ground stations are located in New York City and Berlin. The black vertical lines depict the the maximum secure BSM rate for each ground station pass and demonstrate that a year-round connection is possible with a single repeater constellation. The dashed black curves represent the envelope of all possible connections, including daytime and nighttime passes. The red solid curve shows the accumulated secure BSMs when only nighttime passes are considered. The red dashed curve gives the combined number of daytime and nighttime passes. A total of 51224 secure BSMs are performed on the central satellite, of which 48992 occur during nighttime.}\label{BSM_DL_NB_sso}
\end{figure}

The second orbit configuration can overcome this drawback by using SSOs. Here, the time at which a ground station pass is to take place can be set to nighttime, with the disadvantage that the inter-satellite distances vary during an orbit. \Cref{BSM_DL_NB_sso} demonstrates that with a single repeater constellation, a year-round connection can be achieved with at least one connection per day in fall and winter and at least three times per week during spring and summer. The maximum secure BSM rate is \SI{3.01}{\hertz} similar to the first orbital configuration. Over the course of the year, the combined number of  BSMs during the day and night increases linearly to 51224. As expected, the majority of BSMs are conducted during the night, amounting to 48992.

\section{Discussion and Conclusion}\label{sec:dis}

In this work, we have performed a detailed simulation of the satellite and link dynamics and their impact on the performance of QR constellations in realistic operating scenarios. While previous works have limited their analysis to static links~\cite{Guendogan2021} or to specific reference passes~\cite{fittipaldi2024entanglement}, we have developed an integrated simulator that allows us to expediently configure a satellite constellation and compute the BSM rate for extended periods of time, e.g.\ for determining the annual rates of the repeater link. This is significant as the amount and quality of links that can be established by the repeater constellation varies drastically over time and the system performance cannot be reliably inferred from a single satellite pass. 

The simulator calculates the rotation of Earth and the position of the Sun, which allows us to distinguish between nighttime and daytime passes. It includes the most important perturbation, that is the $J_2$ coefficient related to Earth oblateness. This enables us to comprehensively analyze how sensitive the performance is to variations in orbital configuration, including satellite altitude and inter-satellite distances. The simulator incorporates analytical expressions for the instantaneous BSM rates calculated at each time step. This allows the efficient computation of the optimal storage cutoff times in the presence of losses and errors in the QM, so that performance of the repeater can be assessed over long periods of time. This is in contrast to simulators employing Monte-Carlo based approaches, which are more computationally intensive. It allows for a flexible selection of ground station locations and satellite orbits, rendering it an ideal tool for the design phase of satellite-based QRs.

We have considered downlink and uplink repeater architectures, which have starkly different technological requirements. Our findings show that the downlink architecture has a higher performance compared to the uplink architecture for the assumed parameters. This is not only due to the higher channel efficiencies but, remarkably, also due to the significantly shorter round-trip time of the classical communication, leading to lower requirements for the storage time. Although the downlink architecture exhibits higher rates, unlike the uplink architecture, it requires a QND measurement to herald the presence of a photon in the memory input. Therefore, the choice of architecture is closely linked to the availability of the corresponding hardware.

Furthermore, we have investigated a European (Madrid and Berlin) and an intercontinental scenario (New York City and Berlin) to analyze the operability for ground stations at different distances. The study of the EU scenario revealed that higher BSM rates can be achieved if the outer satellites are placed very close to the central one. This would eliminate the need for a repeater chain of three satellites as long as there is a direct line-of-sight between the central satellite and each ground station. In the IC scenario, the inter-satellite channel is the main source of loss due to the large link distance, resulting in a low annual number of BSMs. 

Our results suggest that further technological improvements compared to the assumptions made in this work are required to make this operating scenario feasible for single-node quantum repeaters. Indeed, lowering the technological assumptions for the quantum memory or the entangled photon source would lead to significantly lower rates. A memory coherence time that is less than the round-trip time of the DL (UL) architecture of \SI{8.1}{\milli\second} (\SI{22.8}{\milli\second}) would lead to rates approaching zero. A reduction in the source repetition rate results in a linear decrease in the secure BSMs. In our IC scenario, a rate below \SI{30}{\mega\hertz} would give secure BSM rate of less than one BSM per second at zenith (DL architecture at \SI{500}{\kilo\meter}). Furthermore, the implementation of a telescope with a diameter of \SI{0.5}{\meter} on a LEO satellite would further increase the complexity and cost. Therefore, compromises have to be made on some technological components to keep costs from getting out of control. However, it should be emphasised that alternative terrestrial solutions would require not only a huge number of intermediate nodes in the case of a fiber-based repeater network, but also the bridging of an ocean. This renders such options likely impractical. 

Lastly, we analyzed two orbital configurations, one based on SSOs and one featuring constant inter-satellite distances. The results show that the former allows to accumulate a higher number of BSM in the long run, as it has nighttime connections throughout the year, while the latter exhibits weeks-long connection gaps. However, if links can also be established during daytime, the latter orbit configuration provides the higher total number of BSMs.

Our quantitative analysis indicates that future work should extend the architecture to a constellation of five satellites realizing a two-node-repeater. This allows shorter inter-satellite distances and thus lower channel losses in the individual links. For ideal QMs with heralding, this results in a maximum expected BSM rate of the order $\eta^{1/3}$, where $\eta$ is the total channel efficiency over the entire chain. However, in the presence of imperfections such as limited heralding and memory coupling efficiencies, additional nodes do not necessarily increase the performance and therefore the optimal number of nodes has to be determined individually for each scenario.

Another natural extension would be to consider multi-user networks. From an operational point of view, it would be economical if the same QR constellation were able to provide multiple ground stations with entangled photons. Our analysis shows that the inter-satellite distance, which is determined by the angular separation of the ground stations, strongly influences the QR performance. Specifically for the IC scenario, shortening the inter-satellite distance compared to the optimal distance leads to a steep reduction in the number of secure BSMs. Therefore, the challenge is to design a satellite constellation that can provide multiple sets of ground stations with entangled photons. One could also imagine strategies to circumvent bad local weather conditions by flexibly connecting to ground stations where the link is not obstructed by clouds.

Our results indicate that at large distances to the end-user, lower orbits lead to higher system performance. The biggest disturbance force for LEO satellites below an altitude of \SI{500}{\kilo\meter} is atmospheric drag. This effect has not been taken into account so far, but would severely limit the lifetime of the satellites and should be considered when evaluating lower orbits.

In conclusion, we have developed a simulator that allows us to efficiently perform an in-depth analysis of one-node three-satellite QR constellations over long time periods, rendering it the perfect tool for both theoretical studies and mission design.
\backmatter

\bmhead{Acknowledgements}
This version of the article has been accepted for publication, after peer review but is not the Version of Record and does not reflect post-acceptance improvements, or any corrections. The Version of Record is available online at:
\href{https://doi.org/10.1140/epjqt/s40507-025-00307-8}{https://doi.org/10.1140/epjqt/s40507-025-00307-8}

\bmhead{Funding}
The work on which this report is based was done within the project QuNET funded by the German Federal Ministry of Education and Research under the funding code 16KIS1265. The authors are responsible for the content of this publication.

\bmhead{Abbreviations}          
AO, Adaptive Optics;
AOCS, Attitude and Orbit Control System;
BSM, Bell State Measurement;
DL, Downlink;
EPS, Entangled Photon Source;
EU, Europe;
GS, Ground Station;
HV, Hufnagel-Valley;
IC, Intercontinental;
IS, Inter-Satellite;
LEO, Low Earth Orbit;
LGS, Laser Guide Star;
LoS, Line-of-Sight;
QBER, Quantum Bit Error Rate;
QKD, Quantum Key Distribution;
QM, Quantum Memory;
QND, Quantum Non-Demolition;
QR, Quantum Repeater;
RAAN, Right Ascension of the Ascending Node;
SMF, Single-Mode Fiber;
SPD, Single-Photon Detector;
SSO, Sun-Synchronous Orbit;
UL, Uplink;

\section*{Declarations}

\bmhead{Availability of data and materials}
Data sharing is not applicable to this article as no datasets were generated or analysed during the current study.

\bmhead{Competing interests}
We declare that Dr. Wörner is the head of the Institute for Satellite Geodesy and Inertial Sensing at DLR. The authors declare that they have no other competing interests.

\bmhead{Authors' contribution}
J.M. designed the overall simulation model, carried out the orbit simulations and implemented the optical link model. P.K. derived the analytical formulas to describe the entanglement swapping process. D.O. derived the formula for the collection efficiency and advised on the optical link models. All authors read and approved the final manuscript. 

\begin{appendices}

\section{Entanglement swapping scheme}\label{sec:A1}
\paragraph{Attempted BSMs}
    As described in the main text, a BSM will be attempted whenever both links succeed within a certain cutoff time $t_\upA^\upc$ or $t_\upB^\upc$, depending on which link succeeds first. The protocol is assumed to be performed in a pulsed manner, i.e. consisting of discrete, consecutive trials to generate entanglement in the corresponding links with a repetition rate $R$. Thus, it is convenient to express the time $t$ in terms of a number of discrete time-bins $t = d/R$. 
    We can derive the rate of such attempts by looking at the probability of performing a BSM between a qubit for which the classical information has just been received and an earlier qubit in the other memory. This is given by the probability that there was an event in the other memory within the cutoff time, multiplied by the probability that this has not already been paired. The following system of equations is obtained
    \begin{subequations}
        \begin{align}
            P_{\BSM|\upA} &= P(\upB, d_\upB) [ 1 - P_{\BSM|\upB} ] \,,\\
            P_{\BSM|\upB} &= P(\upA, d_\upA) [ 1 - P_{\BSM|\upA} ] \,,
        \end{align}
    \end{subequations}
    whose solution reads
    \begin{subequations}
        \begin{align}
            P_{\BSM|\upA} &= \frac{\bar{P}(\upA,d_\upA) P(\upB,d_\upB)}{1 - P(\upA,d_\upA) P(\upB,d_\upB)} \,,\\
            P_{\BSM|\upB} &= \frac{P(\upA,d_\upA) \bar{P}(\upB,d_\upB)}{1 - P(\upA,d_\upA) P(\upB,d_\upB)} \,,
        \end{align}
    \end{subequations}
    where the bar represents the complementary probability, i.e. $\bar{x} = 1 - x$.
    Due to the discrete time-bins, the corresponding probabilities are given by a geometric distribution
    \begin{subequations}
        \begin{align}
            P(\upA,d_\upA) &= \eta_\upA \bar{\eta}_\upB \sum_{d=1}^{d_\upA} (\bar{\eta}_\upA \bar{\eta}_\upB)^{d-1} = \frac{1 - (\bar{\eta}_\upA \bar{\eta}_\upB)^{d_\upA}}{1 - \bar{\eta}_\upA \bar{\eta}_\upB} \eta_\upA \bar{\eta}_\upB \,,\\
            P(\upB,d_\upB) &= \bar{\eta}_\upA \eta_\upB \sum_{d=1}^{d_\upB} (\bar{\eta}_\upA \bar{\eta}_\upB)^{d-1} = \frac{1 - (\bar{\eta}_\upA \bar{\eta}_\upB)^{d_\upB}}{1 - \bar{\eta}_\upA \bar{\eta}_\upB} \bar{\eta}_\upA \eta_\upB \,.
        \end{align}
    \end{subequations}
    Here, an event in one of the memories always refers to the time at which the classical information about its entangled partner was received. Therefore, the effective cutoff time $d_\upA, d_\upB$, that is the maximum number of time-bins that a qubit will be stored after an event, is given by the maximum storage time of the actual qubit, corrected by the round-trip time of the classical information: $d_\upA = d_\upA^\upc - d_\upA^\circlearrowleft = R (t_\upA^\upc - t_\upA^\circlearrowleft)$ and $d_\upB = d_\upB^\upc - d_\upB^\circlearrowleft = R (t_\upB^\upc - t_\upB^\circlearrowleft)$. To make any meaningful use of the memories, each qubit needs to be stored at least until the classical information about its entangled partner has arrived, so that the effective cutoff times are positive.
    
    The total number of attempted BSMs is obtained by weighing these expressions by their expected number of occurrences
    \begin{equation}
        N_{\BSM} = N_{\upA\upB} P_{\BSM|\upA\upB} + N_\upA P_{\BSM|\upA} + N_\upB P_{\BSM|\upB} \,,
    \end{equation}
    where $N_{\upA\upB} = N \eta_\upA \eta_\upB$, $N_\upA = N \eta_\upA \bar{\eta}_\upB$ and $N_\upB = N \bar{\eta}_\upA \eta_\upB$, with $N$ being the total number of trials under consideration. The probability of attempting a BSM when both events occur simultaneously is simply $P_{\BSM|\upA\upB} = 1$.
    
\paragraph{Successful BSMs}
    The exponential memory losses can be expressed in terms of discrete time-bins according to
    \begin{subequations}
    \begin{align}
            P(\checkmark|\upA, d) &= \eta_\upA^\checkmark \upe^{-t/\tau_\upA} = \eta_\upA^\checkmark p_\upA^d \,,\\
            P(\checkmark|\upB, d) &= \eta_\upB^\checkmark \upe^{-t/\tau_\upB} = \eta_\upB^\checkmark p_\upB^d \,,
        \end{align}
    \end{subequations}
    where $p = \upe^{-1/(\tau R)}$. In a similar approach to above, the probabilities that some event can be used for a successful BSM can be derived, that is a BSM where both qubits are still retrieved from the memory and hence a valid result is obtained. As both qubits of a pair will always be stored in the memory for a minimum of the corresponding round-trip time the resulting probabilities are given by
    \begin{subequations}
        \begin{align}
            P_{\BSM|\upA}^\checkmark &= P(\upB,d_\upB,\checkmark) P(\checkmark|\upA,d_\upA^\circlearrowleft) [ 1 - P_{\BSM|\upB} ] \,,\\
            P_{\BSM|\upB}^\checkmark &= P(\upA,d_\upA,\checkmark) P(\checkmark|\upB,d_\upB^\circlearrowleft) [ 1 - P_{\BSM|\upA} ] \,,
        \end{align}
    \end{subequations}
    where
    \begin{subequations}
        \begin{align}
            P(\upA,d_\upA,\checkmark) &= \eta_\upA^\checkmark \eta_\upA \bar{\eta}_\upB p_\upA^{d_\upA^\circlearrowleft} \sum_{d=1}^{d_\upA} (\bar{\eta}_\upA \bar{\eta}_\upB p_\upA)^{d-1} = \frac{1 - (\bar{\eta}_\upA \bar{\eta}_\upB p_\upA)^{d_\upA}}{1 - \bar{\eta}_\upA \bar{\eta}_\upB p_\upA} \eta_\upA \bar{\eta}_\upB \eta_\upA^\checkmark p_\upA^{d_\upA^\circlearrowleft} \,,\\
            P(\upB,d_\upB,\checkmark) &= \eta_\upB^\checkmark \bar{\eta}_\upA \eta_\upB p_\upB^{d_\upB^\circlearrowleft} \sum_{d=1}^{d_\upB} (\bar{\eta}_\upA \bar{\eta}_\upB p_\upB)^{d-1} = \frac{1 - (\bar{\eta}_\upA \bar{\eta}_\upB p_\upB)^{d_\upB}}{1 - \bar{\eta}_\upA \bar{\eta}_\upB p_\upB} \bar{\eta}_\upA \eta_\upB \eta_\upB^\checkmark p_\upB^{d_\upB^\circlearrowleft} \,.
        \end{align}
    \end{subequations}
    Again, the total number of such events is obtained by weighing these expressions by their expected number of occurrences
    \begin{equation}
        N_\BSM^\checkmark = N_{\upA\upB} P_{\BSM|\upA\upB}^\checkmark + N_\upA P_{\BSM|\upA}^\checkmark + N_\upB P_{\BSM|\upB}^\checkmark \,,
    \end{equation}
    where $P_{\BSM|\upA\upB}^\checkmark = P(\checkmark|\upA,d_\upA^\circlearrowleft) P(\checkmark|\upB,d_\upB^\circlearrowleft)$.

\paragraph{Correct BSMs}
    In analogy to the losses, the exponential flipping rates can be expressed in terms of the discrete time-bins as
    \begin{subequations}
        \begin{align}
            P(\pm|\upA, d) &= \frac{1 \pm \eta_\upA^+ (\mathcal{P}_\upA)^d}{2} \,,\\
            P(\pm|\upB, d) &= \frac{1 \pm \eta_\upB^+ (\mathcal{P}_\upB)^d}{2} \,,
        \end{align}
    \end{subequations}
    where $\mathcal{P} = \upe^{-1/(\mathcal{T} R)}$ and $\eta_\upA^+, \eta_\upB^+$ are the probabilities of a flip occurring on read-in or read-out. The following system of equations is obtained
    \begin{subequations}
        \begin{align}
            P_{\BSM|\upA}^\pm &= P(\upB,d_\upB,\pm|\upA,d_\upA^\circlearrowleft) [ 1 - P_{\BSM|\upB} ] \,,\\
            P_{\BSM|\upB}^\pm &= P(\upA,d_\upA,\pm|\upB,d_\upB^\circlearrowleft) [ 1 - P_{\BSM|\upA} ] \,.
        \end{align}
    \end{subequations}
    Due to the fact that two flips correspond to a correct result again, the corresponding probabilities are given by
    \begin{subequations}
        \begin{align}
            P(\upA,d_\upA,\pm|\upB,d_\upB^\circlearrowleft) &= \eta_\upA^\checkmark \eta_\upA \bar{\eta}_\upB p_\upA^{d_\upA^\circlearrowleft} \sum_{d=1}^{d_\upA} (\bar{\eta}_\upA \bar{\eta}_\upB p_\upA)^{d-1} P(+|d, d_\upB^\circlearrowleft) \nonumber\\
            &= \frac{1}{2} P(\upA,d_\upA,\checkmark) \pm \frac{1}{2} \Delta P(\upA,d_\upA,+|\upB,d_\upB^\circlearrowleft) \,,\\
            P(\upB,d_\upB,\pm|\upA,d_\upA^\circlearrowleft) &= \eta_\upB^\checkmark \bar{\eta}_\upA \eta_\upB p_\upB^{d_\upB^\circlearrowleft} \sum_{d=1}^{d_\upB} (\bar{\eta}_\upA \bar{\eta}_\upB p_\upB)^{d-1} P(+|d_\upA^\circlearrowleft, d) \nonumber\\
            &= \frac{1}{2} P(\upB,d_\upB,\checkmark) \pm \frac{1}{2} \Delta P(\upB,d_\upB,+|\upA,d_\upA^\circlearrowleft) \,,
        \end{align}
    \end{subequations}
    where
    \begin{subequations}
        \begin{align}
            \Delta P(\upA,d_\upA,+|\upB,d_\upB^\circlearrowleft) &= \frac{1 - (\bar{\eta}_\upA \bar{\eta}_\upB p_\upA \mathcal{P}_\upA)^{d_\upA}}{1 - \bar{\eta}_\upA \bar{\eta}_\upB p_\upA \mathcal{P}_\upA} \eta_\upA \bar{\eta}_\upB \eta_\upA^\checkmark \eta_\upA^+ \eta_\upB^+ (p_\upA \mathcal{P}_\upA)^{d_\upA^\circlearrowleft} (\mathcal{P}_\upB)^{d_\upB^\circlearrowleft} \,,\\
            \Delta P(\upB,d_\upB,+|\upA,d_\upA^\circlearrowleft) &= \frac{1 - (\bar{\eta}_\upA \bar{\eta}_\upB p_\upB \mathcal{P}_\upB)^{d_\upB}}{1 - \bar{\eta}_\upA \bar{\eta}_\upB p_\upB \mathcal{P}_\upB} \bar{\eta}_\upA \eta_\upB \eta_\upB^\checkmark \eta_\upA^+ \eta_\upB^+ (p_\upB \mathcal{P}_\upB)^{d_\upB^\circlearrowleft} (\mathcal{P}_\upA)^{d_\upA^\circlearrowleft} \,.
        \end{align}
    \end{subequations}
    This leads to 
    \begin{equation}
        N_{\BSM}^\pm = N_{\upA\upB} P_{\BSM|\upA\upB}^\pm + N_\upA P_{\BSM|\upA}^\pm + N_\upB P_{\BSM|\upB}^\pm = \frac{1}{2} N_{\BSM}^\checkmark \pm \frac{1}{2} \Delta N_\BSM^+ \,,
    \end{equation}
    where $P_{\BSM|\upA\upB}^\pm = P(\checkmark|\upA,d_\upA^\circlearrowleft) P(\checkmark|\upB,d_\upB^\circlearrowleft) P(\pm|d_\upA^\circlearrowleft, d_\upB^\circlearrowleft)$. Thus, the correct/erroneous BSMs can be composed by a contribution $N_\BSM^\checkmark/2$, corresponding to uncorrelated events, corrected by the number of correlated BSMs
    \begin{equation}
        \Delta N_\BSM^+ = N_{\upA\upB} \Delta P_{\BSM|\upA\upB}^+ + N_\upA \Delta P_{\BSM|\upA}^+ + N_\upB P_{\BSM|\upB}^+ \,.
    \end{equation}
    Here, $\Delta P_{\BSM|\upA}^+$ and $\Delta P_{\BSM|\upB}^+$ are given by
    \begin{subequations}
        \begin{align}
            \Delta P_{\BSM|\upA}^+ &= \Delta P(\upB, d_\upB, +|\upA, d_\upA^\circlearrowleft) [1 - P_{\BSM|\upB}],\\
            \Delta P_{\BSM|\upB}^+ &= \Delta P(\upA, d_\upA, +|\upB, d_\upB^\circlearrowleft) [1 - P_{\BSM|\upA}]
        \end{align}
    \end{subequations}
    and $\Delta P_{\BSM|\upA\upB}^+ = P(\checkmark|\upA, d_\upA^\circlearrowleft) P(\checkmark|\upB, d_\upB^\circlearrowleft) \eta_\upA^+ \eta_\upB^+ \upe^{-t_\upA/\mathcal{T}_\upA} \upe^{-t_\upB/\mathcal{T}_\upB}$.

\section{Channel efficiency}\label{sec:A2}

\subsection{Derivation of the correction factor of the collection efficiency } \label{sec:coll}

Here, we derive the correction factor to compute the collection efficiency in the near-field case, that is, when the transversal size of the beam at the receiver is comparable to the receiver diameter. We give an analytical expression for the case where the beam shape is Gaussian (with beam waist $w$ at the receiver) and the aperture is circular (with diameter $D_\Rx$). Using polar coordinates $(r,\phi)$, the intensity pattern of the beam is
\begin{align}
    I(r,\phi) = I_\text{peak} r \upe^{-2\frac{r^2}{w^2}},
\end{align}
where the total transmitted power is 
\begin{align}
    P_\Tx = (\uppi/2) I_\text{peak} w^2
    \quad \Longleftrightarrow \quad
    w^2 = \frac{P_\Tx}{(\uppi/2) I_\text{peak}} . \label{eq:w}
\end{align}
The collection efficiency is then given by
\begin{align}
	\eta_{\text{coll}} 
	& = 
	\frac{1}{P_\Tx}\int_0^{2\uppi} \int_0^{D_\Rx/2} 
	I_\text{peak} r \upe^{-2\frac{r^2}{w^2}} \, \upd r\, \upd\phi \nonumber\\
	& = 
	\frac{(\uppi/2) I_\text{peak} w^2}{P_\Tx} 
	\left(1 - \upe^{-\frac{D_\Rx^2}{2w^2}}\right)  
	\nonumber\\
	& = 
	1-\upe^{-\eta_{\text{coll,ff}}},
\end{align}
where we have used that in the far-field the collection efficiency is 
\begin{align}
    \eta_{\text{coll,ff}} = 
    \frac{A_\Rx I_\text{peak}}{P_\Tx} =
    \frac{(4/\uppi)A_\Rx}{2 \frac{P_\Tx}{(\uppi/2)I_\text{peak}}} =
    \frac{D_\Rx^2}{2w^2} .
\end{align}

\subsection{SMF coupling efficiency} \label{sec:SMF}

The SMF coupling efficiency for the downlink signal can be divided into two independent components
\begin{align}
    \eta_{\text{smf}} = \eta_{0}\eta_{\text{ao}}. 
\end{align}
The optical coupling efficiency $\eta_0$ of the receiver describes the efficiency of coupling the unperturbed free-space beam into a SMF. For a telescope with a uniformly illuminated circular aperture and SMF transmitting a Gaussian mode, $\eta_0$ can be approximated as $\eta_0 \approx 81\%$, corresponding  to a loss of $\SI{0.89}{dB}$~\cite{Scriminich2022}.

Due to atmospheric turbulence in the lower part of the atmosphere, the signal accumulates wavefront distortions. The strength of the turbulence is expressed by the refractive-index structure constant $C_n^2(h)$, which we model with a Hufnagel-Valley (HV) turbulence profile~\cite{Valley1980, WolfeWilliamL1989Tih, Hardy1998}: 
\begin{equation}
	C_n^2(h) = A\exp\bigg(-\frac{h}{H_A}\bigg) + B\exp\bigg(-\frac{h}{H_B}\bigg) + Ch^{10}\exp\bigg(-\frac{h}{H_C}\bigg).
\end{equation}
In this generalized HV-model, $A$ defines the boundary layer turbulence strength, $H_A$ the height of the $1/\text{e}$ decay, while $B$ and $H_B$ define the turbulence in the troposphere up to \SI{10}{\kilo\meter} and $C$ and $H_C$ the turbulence peak at the tropopause at \SI{10}{\kilo\meter}. Values for different turbulence profiles can be found in Ref.~\cite{Pugh2020}.

The Fried parameter $r_0$ describes the atmospheric coherence width (i.e., the characteristic length of wavefront perturbations) and is given by~\cite{Fried1965, Parenti1994, Stotts2017}
\begin{equation}
	r_0 = \bigg(\frac{0.423k^2}{|\sin{\theta}|}\int_0^H C_n^2(h)\mathrm{d}h\bigg)^{-3/5},
\end{equation}
with the wavenumber $k = 2\pi/\lambda$, the elevation angle $\theta$, the refractive-index structure constant $C_n^2(h)$ and the satellite altitude $H$. As the Fried parameter depends on the elevation angle, the fiber coupling efficiency changes over the course of the satellite pass, taking the lowest value for small elevation angles.

The efficiency after correcting these wavefront perturbations by applying adaptive optics $\eta_{\text{ao}}$ can be expressed as~\cite{Ma2015, Canuet2017, Scriminich2022} 
\begin{equation}\label{eq::SMF_AO}
	\langle \eta_{\text{ao}} \rangle = \prod_{\substack{n,m\\n>n_{\max}}}\frac{1}{\sqrt{1+2\langle (b_n^m)^2\rangle}},
\end{equation} 
where $b_n^m$ are the expansion coefficients of the wavefront in terms of Zernike polynomials ($n$ is the radial order and $m$ is the azimuthal order). This equation then means that the wavefront errors can be exactly suppressed up to the radial order $n\leq n_{\text{max}}$. This requires a deformable mirror with approximately $n_{\text{max}}^2$ actuators and assumes that the speed of the AO control loop is much faster than the atmospheric channel coherence time~\cite{Scriminich2022}. The zeroth order represents a global phase shift, the first order fluctuations in the angle of attack and $n>1$ higher order wave front errors~\cite{Noll1976}. The Zernike coefficients $b_n^m$ are modelled as independent and identically distributed random variables based on Gaussian statistics that have no dependence on $m$. Their variance $\langle (b_n^m)^2\rangle$ is given by~\cite{Noll1976, Boreman1996}
\begin{equation}
	\langle (b_n^m)^2\rangle = \bigg(\frac{D_{\text{Rx}}}{r_0}\bigg)^{\frac{5}{3}}\frac{n+1}{\uppi}\frac{\Gamma(n-\frac{5}{6})\Gamma(\frac{23}{6})\Gamma(\frac{11}{6})\sin{\frac{5}{6}\uppi}}{\Gamma(n+\frac{23}{6})}, \label{eq:bnm}
\end{equation}
where $\Gamma$ is Euler's gamma function and $D_{\text{Rx}}$ is the receiver diameter.

\subsection{Beam wandering and higher-order wavefront distortions}\label{sec:bwb}

Beam wandering and higher-order wavefront distortions (beam broadening) lead to losses for the uplink signal. Adaptive optics systems with an LGS can partially correct these errors. The resulting link efficiency, including beam wandering, higher wavefront errors and their respective corrections, is calculated according to the model of ref.~\cite{Pugh2020}:
\begin{equation}
	\eta_{\text{bwb}} = 
	S \, \eta_{\text{bw,diff}} + (1-S) \left(\frac{w_\text{diff}}{w_{\text{ST}}}\right)^{\!2} \eta_{\text{bw,ST}}.
\end{equation}
Here, $S$ is the Strehl ratio, $\eta_{\text{bw,diff}}$ and $\eta_{\text{bw,ST}}$ are efficiency terms due to beam wander for a diffraction-limited beam and a short-term turbulence-broadened beam respectively. The beam wander can be partially compensated by tip-tilt corrections, bringing $\eta_{\text{bw,diff}}$ and $\eta_{\text{bw,ST}}$ closer to one. Note that $\eta_{\text{bw,diff}}$ and $\eta_{\text{bw,ST}}$ correspond to $I_\text{diff}$ and $I_\text{ST}$ in ref.~\cite{Pugh2020} and $\eta_{\text{bwb}}$ does not include the collection efficiency. These factors can be computed as described below.

The diffraction-limited beamwidth, evaluated at the satellite in distance $L$, is 
\begin{align}
    w_{\text{diff}}(z=L) = \sqrt{w_0^2\bigg(1+\frac{L^2}{z_0^2}\bigg)}
\end{align}
and the Rayleigh length is $z_0 = \uppi w_0^2/\lambda$, where $w_0$ is the beam waist at the transmitter and $\lambda$ the wavelength. If the tip-tilt error induced by the atmosphere is partially corrected by a fine tracking system at the transmitter, we can define the short-term beamwidth broadened by the residual beam wander as~\cite{Yura1973}:
\begin{equation}
	w_{\text{ST}}(z=L) = \sqrt{w_0^2\bigg(1+\frac{L^2}{z_0^2}\bigg) + 2\bigg(\frac{4.2L}{kr_0}\bigg[1-0.26\big(\frac{r_0}{w_0}\big)^{1/3}\bigg]\bigg)^2},
\end{equation}
where $k=2\pi/\lambda$ is the wave number. The Fried parameter $r_0$ for a spherical wave (which is a valid approximation for uplink beams) is defined as~\cite{Fried1965, Tyson2015}
\begin{equation}
	r_0 = \bigg(\frac{0.423k^2}{|\sin{\theta}|}\int_0^H C_n^2(h)\bigg(1-\frac{h}{H}\bigg)^{5/3}\mathrm{d}h\bigg)^{-3/5},
\end{equation}
with the satellite altitude $H$ and elevation angle $\theta$. In general, assuming a Gaussian beam and a Gaussian-distributed pointing jitter, the beam wander efficiency factor $\eta_{\text{bw}}$ is given by~\cite{Jeganathan1996}
\begin{align}
    \eta_{\text{bw}} = \frac{\beta}{\beta+1} \quad 
    \text{with} \ \beta = \frac{1}{8} \left(\frac{\vartheta}{\sigma}\right)^2,   
    \label{eq:eta_bw}
\end{align}
where $\beta$ is the pointing accuracy relative to the beam divergence, $\sigma$ is standard deviation of the one-dimensional residual beam wander and $\vartheta$ is the full-angle beam divergence. Then we can use~\cref{eq:eta_bw} to compute $\eta_\text{bw,diff}$ and $\eta_\text{bw,ST}$ by inserting the values for $\vartheta_\text{diff}$ and $\vartheta_\text{ST}$, which are given by $\vartheta_\text{diff} = w_\text{diff}/L$ and by $\vartheta_\text{ST} = w_{\text{ST}}/L$ (in the limit $L \gg z_0$). To calculate the variance of the beam wander, we consider four different effects:
\begin{equation}
	\sigma = \sqrt{\sigma^2_{\text{TFD}} + \sigma^2_{\text{CA}} + \sigma^2_{\text{TA}} + \sigma^2_{\text{SNR}}}.
\end{equation} 
The closed-loop tilt feedback delay $\sigma_{\text{TFD}}$ is caused by the finite speed of the beam correction system, $\sigma_{\text{CA}}$ is the estimation error of the centroid measurement, $\sigma_{\text{TA}}$ is the tilt anisoplanatic error, which describes the difference between the measured tilt of the downlink beacon and the tilt of the actual signal, and  $\sigma_{\text{SNR}}$ is the signal-to-noise ratio error. See~\cite{Olivier1993TiptiltC, Tyson2015,  Pugh2020} for further details.

The Strehl ratio is a measure of the quality of the focusing of the signal within the receiver and can be computed as~\cite{Mahajan1983}
\begin{align}
    S=\exp(-\zeta^2),
\end{align}
where $\zeta$ is the standard deviation of the wavefront distortions. The Strehl ratio can be decomposed into three main components~\cite{Tyson2015, Pugh2020}: 
\begin{equation}
    \zeta = \sqrt{\zeta^2_{\text{AFD}}+\zeta^2_{\text{fit}}+\zeta^2_{\text{cone}}}.
\end{equation}
Here, the AO feedback delay error $\zeta_{\text{AFD}}$ is caused by differences in the turbulence conditions at the time the wavefront errors are measured and at the time the corrections are applied, the spatial fitting error $\zeta_{\text{fit}}$ is due to the correction of only a limited number of Zernike polynomials $Z_{\text{max}}$, and the cone effect or focal anisoplanatism $\zeta^2_{\text{cone}}$ is due to the altitude difference between the satellite and the LGS. Specifically, we assume that by using a time-gating camera to measure Rayleigh backscatter of a laser pulse, an artificial star can be created to sample the atmospheric turbulence at an altitude of $H_{\text{LGS}} = \SI{18}{\kilo\meter}$.

Taken together, all these effects result in larger wavefront errors and larger beam wander at lower elevation angles and at higher satellite velocities. 

\section{Simulation parameters} \label{sec:Param}

\begin{table}[ht]
\caption{Communication link parameters}\label{tab::Com}
\begin{tabular*}{\linewidth}{@{\extracolsep{\fill}}lcc}
\toprule
Parameter & Symbol  & Value\\
\midrule
Minimum elevation angle & $\theta_{\mathrm{min}}$   & $\SI{20}{\degree}$  \\
Number of Zernike polynomials corrected UL   & $Z_{\mathrm{max}}$   & $36$ \\
Max order adaptive optics corrections DL   & $n_{\mathrm{max}}$   & $12$  \\
Ground station telescope diameter & $D_{\mathrm{GS}}$   &    $\SI{1}{\meter}$\\
Satellite telescope diameter  & $D_{\mathrm{Sat}}$   &    $\SI{0.5}{\meter}$\\
Correction bandwidth     & $f_c$   & $\SI{50}{\hertz}$ \\
Atmospheric visibility & & \SI{23}{\kilo\meter}\\
Altitude Laser guide star & $H_{\text{LGS}}$ & \SI{18}{\kilo\meter}\\
Beam waist UL & $\omega_{0,\text{UL}}$ & $\SI{0.15}{\meter}$\\
Beam waist DL& $\omega_{0,\text{DL}}$ & $\SI{0.22}{\meter}$\\
Wavelength    & $\lambda$   & $\SI{1550}{\nano\meter}$  \\
Hufnagel-Valley turbulence profile    &    & HV 10-10  \\
Bufton wind model & ($v_{\text{g}}$, $v_{\text{t}}$, $h_{\text{peak}}$, $h_{\text{scale}}$) & ($\SI{5}{\meter\per\second}$, $\SI{20}{\meter\per\second}$, $\SI{9.4}{\kilo\meter}$, $\SI{4.8}{\kilo\meter}$)\\
\botrule
\end{tabular*}
\end{table}

\Cref{tab::Com} summarizes the parameters of the communication link. The minimum elevation angle between satellite and ground station is $\theta_{\text{min}} = 20^\circ$. The turbulence profile is modeled with the Hufnagel-Valley turbulence profile~\cite{Valley1980, WolfeWilliamL1989Tih} HV\,10-10~\cite{Pugh2020}, which characterizes a ground station location with favourable turbulence conditions. This is justified as we only consider nighttime communication, where temperature fluctuations causing turbulence are less pronounced. To allow for minimal signal attenuation, the communication wavelength is $\lambda = \SI{1550}{\nano\meter}$, which is within a good transparency window of the atmosphere. Furthermore, the atmospheric visibility is set to \SI{23}{\kilo\meter}. It is assumed that the receiver and transmitter terminals are optimal, with telescopes that are not obscured by secondary mirrors and without near-field and defocusing effects. The beam waist of the transmitter terminal of the uplink is set to $\omega_{0,\text{UL}}=\SI{0.15}{\meter}$ and for the downlink terminal to $\omega_{0,\text{DL}}=\SI{0.22}{\meter}$. The maximal order of adaptive optics corrections of the downlink signal is 12. The finite correction bandwidth $f_c$, $Z_{\text{max}}$, the altitude of the Laser guide start $H_{\text{LGS}}$ and the wind model are discussed in ref.~\cite{Pugh2020}.

\begin{table}[ht]
\caption{Quantum repeater parameters}\label{tab::QR}
\begin{tabular*}{\linewidth}{@{\extracolsep{\fill}}lcc}
\toprule
Parameter & Symbol  & Value\\
\midrule
Source repetition rate & $R$   &    $\SI{90}{\mega\hertz}$\\
QM decay time & $\tau$   &    $\SI{100}{\milli\second}$\\
QM coherence time & $\mathcal{T}$   &    $\SI{60}{\milli\second}$\\
QM efficiency (read-in and -out)    & $\eta^\checkmark$   & 0.1  \\
QM fidelity (read-in and -out) & $\eta^+$ & 1\\
Entangled photon source efficiency   & $\eta_{\mathrm{EPS}}$   & 0.2  \\
Single photon detection efficiency   & $\eta_{\mathrm{SPD}}$   & 0.95  \\
Quantum non-demolition measurement efficiency   & $\eta_{\mathrm{QND}}$   & 0.8  \\
Bell state measurement efficiency   & $\eta_{\mathrm{BSM}}$   & 0.5  \\
\botrule
\end{tabular*}
\end{table}

The QM efficiency is the combined read-in and read-out efficiency and is set to $\eta_\upA^\checkmark = \eta_\upB^\checkmark = 0.1$. Similar value ranges were reported by~\cite{Ortu2022} for the storage of photonic time-bin qubits in rare-Earth ion doped crystals. The characteristic decay and coherence time of the memory are $\tau = \SI{100}{ms}$ and $\mathcal{T} = \SI{60}{ms}$. In addition, the DL architecture assumes that a QND measurement can be performed with a success probability of $\eta_{\mathrm{QND}} = \SI{80}{\percent}$ to indicate the presence of a photon inside the QM. The entangled photon source has a repetition rate of $R = \SI{90}{MHz}$ with an efficiency of $\eta_{\mathrm{EPS}} = \SI{20}{\percent}$, while the single-photon detectors operate with an efficiency of $\eta_{\mathrm{SPD}} = \SI{95}{\percent}$. 

\begin{table}[ht]
\caption{Orbit simulation parameters}\label{tab::Orbit}
\begin{tabular*}{\linewidth}{@{\extracolsep{\fill}}lcc}
\toprule
Parameter & Symbol  & Value\\
\midrule
Initial date & $t_{\mathrm{init}}$   & ($\SI{2020}{\year}$, $\SI{1}{\month}$, $\SI{1}{\day}$, $\SI{2}{\hour}$, $\SI{0}{\minute}$, $\SI{0}{\second}$)  \\
Zonal harmonic coefficients max degree & $N_{\mathrm{max}}$   & 2  \\
\midrule\midrule
EU scenario ground stations &   & Madrid, Berlin  \\
Distance between GS & $\nu_{\text{sep,gs}}$ & \SI{16.84}{\degree}\\
\midrule
\textbf{Orbit: const. inter-sat. distances} & & \\
Eccentricity & $(e_1, e_2, e_3)$ & 0\\
RAAN & $(\Omega_1, \Omega_2, \Omega_3)$ & \SI{102.8}{\degree}\\
Inclination & $(i_1, i_2, i_3)$  & \SI{61.8}{\degree}\\
Argument of perigee & $(\omega_1, \omega_2, \omega_3)$ & \SI{0}{\degree}\\
\midrule
\midrule
IC scenario ground stations &   & New York City, Berlin  \\
Distance between GS & $\nu_{\text{sep,gs}}$ & \SI{57.60}{\degree}\\
\midrule
\textbf{Orbit: const. inter-sat. distances} & & \\
Satellite altitude & $(h_1, h_2, h_3)$ & (\SI{500}{\kilo\meter}, \SI{505.44}{\kilo\meter}, \SI{505.39}{\kilo\meter})\\
Eccentricity & $(e_1, e_2, e_3)$ & (0, 8.18e-4, 9.58e-4)\\
RAAN & $(\Omega_1, \Omega_2, \Omega_3)$ & ($28.10^\circ, 28.0894^\circ, 28.0885^\circ$)\\
Inclination & $(i_1, i_2, i_3)$ & $(56.70^\circ, 56.7149^\circ, 56.7147^\circ)$\\
Argument of perigee & $(\omega_1, \omega_2, \omega_3)$ & $(0.0^\circ, 353.6335^\circ, 32.9083^\circ)$\\
True anomaly & $(f_1, f_2, f_3)$ & $(317.0^\circ, 352.1689^\circ, 341.7007^\circ)$\\
\midrule
\textbf{Orbit: SSO} & & \\
Satellite altitude & $(h_1, h_2, h_3)$ & (\SI{500}{\kilo\meter}, \SI{504.31}{\kilo\meter}, \SI{503.85}{\kilo\meter})\\
Eccentricity & $(e_1, e_2, e_3)$ & (0, 17e-4, 4.07e-4)\\
RAAN & $(\Omega_1, \Omega_2, \Omega_3)$ & ($68.50^\circ, 114.1435^\circ, 159.2328^\circ$)\\
Inclination & $(i_1, i_2, i_3)$ & $(97.4055^\circ, 97.3974^\circ, 97.3919^\circ)$\\
Argument of perigee & $(\omega_1, \omega_2, \omega_3)$ & $(0.0^\circ, 176.0524^\circ, 22.0563^\circ)$\\
True anomaly & $(f_1, f_2, f_3)$ & $(308.0^\circ, 145.1230^\circ, 297.7079^\circ)$\\
\botrule
\end{tabular*}
\end{table} 

\Cref{tab::Orbit} summarizes the parameters for the two operating scenarios EU and IC and the two orbit configurations. All simulations start at the same initial date. For simulations dealing with a single ground station pass, a spherical potential of Earth is assumed, while for the simulations examining the annual performance of the repeater, the degree of zonal coefficients is increased to two. For the EU scenario, the orbit parameters \textit{eccentricity}, \textit{RAAN}, \textit{inclination} and \textit{argument of perigee} are the same for all satellites, while the \textit{true anomaly} is chosen for each satellite depending on the desired inter-satellite distance. Furthermore, the satellite altitude is varied between \SIrange[]{500}{1000}{\kilo\meter}. The EU scenario only features the orbit configuration in which the satellites are arranged like as if on a string of pearls, with constant inter-satellite distances. For the IC scenario, an additional orbit configuration based on sun-synchronous orbits is considered.

\end{appendices}

\bibliography{bib}

\end{document}